\documentclass[prd,amstext,amsmath,amssymb,superscriptaddress,floatfix,nofootinbib,notitlepage]{revtex4-1}
\usepackage[hyperref]{xcolor}
\usepackage{graphics}
\usepackage{epsfig}
\usepackage{amsmath}
\usepackage{amssymb}
\usepackage{amsthm}
\usepackage[mathscr]{eucal}
\usepackage{graphicx}
\usepackage{subfigure}
\usepackage{bbm}
\usepackage{ulem}
\usepackage{slashed}
\usepackage{hyperref}
\hypersetup{
  colorlinks, breaklinks,
  linkcolor=blue,
  citecolor=blue,
  linktoc=all
}

\newcommand{\Eq}[1]{{Eq.~({\ref{#1}})}}
\newcommand{\Fig}[1]{{Fig.~{\ref{#1}}}}
\newcommand{\ave}[1]{{\langle{#1}\rangle}}
\newcommand{\bea}{\begin{eqnarray}}
\newcommand{\eea}{\end{eqnarray}}
\newcommand{\beq}{\begin{equation}}
\newcommand{\eeq}{\end{equation}}
\newcommand{\beas}{\begin{eqnarray*}}
\newcommand{\eeas}{\end{eqnarray*}}

\newcommand{\one}{\mathbbm{1}}
\newcommand{\equalover}[1]{{\boldmath{\overset{#1}{=}}}}

\def\p{{\bf p}}

\def\x{{\bf x}}
\def\Ms{\ensuremath{m_\sigma}}

\graphicspath{
{./}
}

\begin{document}

\title{Inhomogeneous phases in the quark-meson model with vacuum
  fluctuations }

\author{Stefano~Carignano}
\affiliation{Department of Physics, The University of Texas at El Paso, USA}

\author{Michael~Buballa}
\affiliation{Theoriezentrum, Institut f\"ur Kernphysik, Technische Universit\"at Darmstadt, Germany}

\author{Bernd-Jochen~Schaefer}
\affiliation{Institut f\"ur Theoretische Physik,
  Justus-Liebig-Universit\"at Gie{\ss}en, Germany}

\begin{abstract}

  Inhomogeneous chiral-symmetry breaking phases at non-vanishing
  chemical potential and temperature are studied within a two-flavor
  quark-meson model in the chiral limit.  The analysis is performed
  beyond the standard mean-field approximation by taking into account
  the Dirac-sea contributions of the quarks.  Compared with the case
  where the Dirac sea is neglected, we find that the inhomogeneous
  phase shrinks, but in general does not disappear.  It is shown
  within a Ginzburg-Landau analysis that the Lifshitz point of the
  inhomogeneous phase coincides with the tricritical point if the
  ratio between sigma-meson and constituent quark mass in vacuum is
  chosen to be $m_\sigma/M = 2$, corresponding to the fixed mass ratio
  in the Nambu--Jona-Lasinio model.  In the present model, however,
  this ratio can be varied, offering the possibility to separate the
  two points.  This is confirmed by our numerical calculations, which
  demonstrate a strong sensitivity of the size of the inhomogeneous
  phase on $m_\sigma$. Finally, we uncover a general instability of
  the model with respect to large wave numbers of the chiral
  modulations, which calls for further improvements beyond the present
  approximation.

\end{abstract}

\maketitle

\section{Introduction}

In spite of the very intensive theoretical and experimental efforts
over the past decades, the phase diagram of Quantum Chromodynamics
(QCD) is still poorly understood \cite{BraunMunzinger:2009zz,
  Fukushima:2010bq}.  Most of our current understanding of the QCD
phase structure revolves around vanishing quark chemical potential,
where ab-initio lattice simulations are reliable \cite{Aoki:2006we,
  Borsanyi:2012vn, Ratti:2013uta, Bazavov:2012kf}.  In the regime of
non-vanishing chemical potential (or net-baryon density), on the other
hand, many fundamental questions are still left unanswered, most
notably on the nature of the chiral phase transition at low
temperatures and the existence of a critical endpoint, for a review
see for example~\cite{Friman:2011zz}.

In the past few years, a growing number of indications has been found
that strong-interaction matter at high densities might form spatially
inhomogeneous phases.  This would lead to a dramatic revision of our
picture of finite-density QCD.  In particular, it has been proposed
that the conjectured critical endpoint (CP) of the first-order chiral
phase transition might be replaced by a so-called Lifshitz point (LP),
where two homogeneous phases and an inhomogeneous phase meet
\cite{Nickel:2009ke}.

The concept of spatially dependent order parameters is rather common
in condensed matter physics, such as the FFLO state in
superconductivity \cite{Fulde:1964zz, larkin:1964zz} or spin density
waves \cite{Overhauser:1962zz}, including magnetism
\cite{PhysRevLett.103.207201}.  It has been discussed for color
superconductivity~\cite{Alford:2000ze, Anglani:2013gfu} and for
nuclear matter in the context of pion
condensation~\cite{Dautry:1979bk}, Skyrme
crystals~\cite{Goldhaber:1987pb}, and recently in an extended linear
sigma model with vector and axial vector mesons \cite{Heinz:2013hza}.
Inhomogeneous chiral-symmetry breaking phases in quark matter have
been proposed already long time ago \cite{Broniowski:1990dy,
  Broniowski:1990gb, Sadzikowski:2000ap, Nakano:2004cd} but received
enhanced attention during the last few years \cite{Nickel:2009wj,
  Carignano:2010ac, Carignano:2011gr, Fukushima:2012mz,
  Carignano:2012sx, Muller:2013tya, Tatsumi:2013nga}.

Most of the current studies on strong-interaction matter at finite
density are performed within QCD-inspired effective models in
mean-field approximation.  One of the most widely employed examples is
the Nambu--Jona-Lasinio (NJL) model, which, in spite of its simple
four-fermion interaction, can be used to describe the phenomenon of
spontaneous chiral symmetry breaking and its restoration at finite
temperature and density~\cite{Vogl:1991qt, Klevansky:1992qe,
  Hatsuda:1994pi, Buballa:2003qv}.  The study of inhomogeneous phases
in the NJL model has been performed by several authors
\cite{Sadzikowski:2000ap, Nakano:2004cd, Nickel:2009wj,
  Carignano:2010ac, Carignano:2011gr, Carignano:2012sx,
  Tatsumi:2013nga}, all with the conclusion that the usual first-order
chiral phase transition for homogeneous quark-matter at low
temperatures and intermediate densities is surrounded by an
inhomogeneous ``island''.  The favored shape for the chiral order
parameter is typically
 one with a one-dimensional spatial modulation  
\cite{Carignano:2012sx, Abuki:2011pf}.

A major drawback of the NJL model is its non-renormalizability: in
order to obtain meaningful results, the diverging vacuum contribution
from the Dirac sea of the quarks to the grand potential has to be
regularized. This introduces ambiguities, since the results depend on
the choice of the regularization prescription as well as the value of
the corresponding cutoff parameter.  In particular, model results are
not expected to be reliable for temperatures and chemical potentials
bigger than the employed ultraviolet cutoff, which is typically chosen
in the range between 0.5 and 1~GeV.
 
A delicate issue arises when one tries to push the NJL model to high
densities, in the regime where large-$N_c$ studies predict the
formation of crystalline phases \cite{Deryagin:1992rw, Shuster:1999tn,
  Kojo:2009ha}.  It was recently found that beyond the inhomogeneous
island a second inhomogeneous phase emerges and seems to persist up to
arbitrarily high chemical potentials. Thorough investigations have
established that this inhomogeneous ``continent'' arises from the
interplay of the medium and the vacuum contributions to the
thermodynamic potential and might at least partially be caused by the
used regularization \cite{Broniowski:1990gb,Carignano:2011gr}.  Since
these findings are not finally clarified, a similar study of
inhomogeneous phases within a renormalizable model is desirable.

For this purpose, and in order to get rid of regularization
ambiguities in general, we consider a two-flavor chiral quark-meson
(QM) model \cite{Scavenius:2000qd, Schaefer:2006ds}, which represents
a renormalizable alternative to the NJL model.  In the first QM-model
studies the vacuum contribution of the quarks to the grand potential
has been neglected because it was expected that a proper refit of the
model parameters would compensate the omission of the Dirac
sea. However, this "no-sea'' or "standard mean-field'' approximation
(sMFA) has been shown to lead to inconsistent thermodynamics, such as
a first-order chiral phase transition in the chiral limit for two
flavors at vanishing chemical potential \cite{Skokov:2010sf}.  In this
context, the importance of the fermionic vacuum fluctuations for the
thermodynamics has been emphasized recently in \cite{Schaefer:2011ex,
  Andersen:2011pr, Gupta:2011ez} and for three flavors in
\cite{Mao:2009aq, Gupta:2009fg, Schaefer:2009ui, Chatterjee:2011jd}.

Inhomogeneous phases in the QM model have already been studied in
Ref.~\cite{Nickel:2009wj}, but again without taking into account the
quark vacuum contributions.  In the present work, we improve these
investigations by going beyond the sMFA and considering a properly
renormalized thermodynamic potential.  In this way we obtain a deeper
understanding of the effect of quantum and thermal fluctuations on the
formation of inhomogeneous phases in QCD matter.

The structure of the paper is as follows: In Sec.~\ref{sec:qmmodel} we
introduce the model Lagrangian and derive a general expression for the
thermodynamic potential, which is the starting point for our later
investigations of the phase structure.  In Sec.~\ref{sec:vac} we
restrict our studies to the homogeneous vacuum, focusing on the
renormalization of the Dirac-sea contributions.  Based on this, we
present numerical results for the phase diagram for homogeneous phases
in Sec.~\ref{sec:pdhom}.  After that our analysis is extended to
include inhomogeneous matter.  We begin with a general Ginzburg-Landau
analysis in Sec.~\ref{sec:GL}, concentrating on the locations of the
CP and the LP, and then explore the phase structure for specific
one-dimensional modulations in Sec.~\ref{sec:pdinhom}.  In
Sec.~\ref{sec:continent} our studies are extended to higher chemical
potentials, analyzing the existence of a second inhomogeneous phase
and its physical nature.  Our conclusions are drawn in
Sec~\ref{sec:conclusions}.

\section{Thermodynamic potential}
\label{sec:qmmodel}

The quark-meson model is defined by the Minkowskian Lagrangian density 
\cite{Scavenius:2000qd,Schaefer:2006ds, Scadron:2013vba, Jungnickel1996b}
\beq
\mathcal{L}_\text{QM}
=
\bar{\psi}
\left(
i\gamma^\mu \partial_\mu
-
g(\sigma+i\gamma_5 \vec\tau \cdot \vec\pi)
\right)
\psi
+
\mathcal{L}_\text{M}^\text{kin}
-
U(\sigma,\vec\pi )
\ ,
\eeq
where $\psi$ is a $4N_f N_c$-dimensional quark spinor with $N_f = 2$
flavor and $N_c = 3$ color degrees of freedom, $\sigma$ is the scalar
field of the sigma meson and $\vec \pi$ the pseudo-scalar fields of the
pion triplet.  The meson kinetic contributions reads 
\beq
\mathcal{L}_\text{M}^\text{kin} = \frac{1}{2} \left(
\partial_\mu\sigma \partial^\mu\sigma 
+
\partial_\mu\vec\pi \partial^\mu\vec\pi
\right) \ ,
\eeq
while
\beq
\label{eq:Uqm}
U(\sigma,\vec\pi)
=
\frac{\lambda}{4}
\left(
\sigma^2
+
\vec\pi^2
-
v^2
\right)^{2}
\,
\eeq
is an $O(4)$-symmetric meson potential.  The $O(4)$-symmetry, which is
isomorphic to the chiral $SU(2)_L\times SU(2)_R$ symmetry, can be
broken explicitly by adding a term, which is linear in the $\sigma$
field.  However, in this work we will restrict ourselves to the chiral
limit and omit such a term. The three model parameters, $g$, $\lambda$
and $v^2$, will be fitted to vacuum properties as discussed in the
next section.

The thermodynamic properties of the model are encoded in the grand
potential per volume $V$,
\bea
\Omega(T,\mu) 
&=&
-\frac{T}{V} \log\mathcal{Z}(T,\mu)
= 
-\frac{T}{V}\log 
\int \mathcal{D}\bar{\psi}\mathcal{D}\psi\mathcal{D}\sigma\mathcal{D}\vec\pi
\exp\left( \int_{V_4} d^4x_E \, (\mathcal{L}_\text{QM}+\mu \bar{\psi}\gamma^0 \psi)\right)\,,
\eea
where $\mathcal{Z}(T,\mu)$ denotes the grand canonical partition
function, which depends on the temperature $T$ and the quark chemical
potential $\mu$.  The integral in the exponent is performed in
Euclidean space-time, $x_E = (\tau, \x)$ with the imaginary time $\tau
= it$, and extends over the four-volume $V_4 = [0,\frac{1}{T}] \times
V$.

Since the Lagrangian is bilinear in the quark fields, the integration
over the quarks can be performed analytically.  In mean-field
approximation we treat the meson fields $\sigma$ and $\pi^{a}$ as
classical and replace them by their expectation
values~\cite{Scavenius:2000qd, Schaefer:2006ds}.  Thereby we also
assume these mean fields to be time independent but, in order to allow
for inhomogeneous phases, we retain their dependence on the spatial
coordinate $\x$.  The thermodynamic potential  then takes the form
\bea
\label{eq:Omega1QM}
\Omega_\text{MFA} (T,\mu;\sigma, \vec{\pi})
&=&
-
\frac{T}{V}
\mathbf{Tr} \, \mathrm{Log}\left( \frac{1}{T}\left(i\partial_{0} - H_\text{QM}+\mu\right)\right)
+
\frac{1}{V}\int_V d^3x\,\mathcal{H}_\text{M}\,,
\eea
where the functional trace runs over $V_4$ and internal (color,
flavor and Dirac) degrees of freedom.\footnote{ We will always
  refer to $\Omega_\text{MFA}$ for any given $\sigma$ and $\vec\pi$ as
  ``thermodynamic potential'' even though strictly speaking this term
  should only be used for the function evaluated at the physical
  minimum.}  The mesonic Hamiltonian density
and the effective Dirac Hamiltonian are given by \beq
\label{eq:Hmes}
       \mathcal{H}_\text{M}
       =
       \frac{1}{2}\left( (\vec\nabla\sigma(\x))^2  + (\vec\nabla
         \vec\pi (\x))^2 \right) 
       +  U(\sigma(\x), \vec\pi (\x))
\eeq
and
\bea
\label{eq:hamiltonian0QM}
H_\text{QM}
&=&
-i\gamma^0\gamma^{i}\partial_{i} +
\gamma^0\left(g\sigma(\x)+ig\gamma^5 \vec \tau \cdot \vec\pi (\x)\right) \,,
\eea
respectively. The latter is hermitean, so that, at least in principle,
it can be diagonalized.  Moreover, since $H_\text{QM}$ is
time-independent, the integration over Euclidean time can be performed
in the usual way by summing fermionic Matsubara frequencies.  One
obtains
\beq
-\frac{T}{V}
\mathbf{Tr} \, \mathrm{Log}\left(\frac{1}{T}\left(i\partial_{0} - H_\text{QM}+\mu\right)\right)
=
-
\frac{1}{V}
\sum_{\lambda} \left[ \frac{E_\lambda -\mu}{2} 
  + T \ln\left(1 + e^{-\frac{E_\lambda - \mu}{T}}\right)\right] \,,
\eeq
where $\lambda$ labels the eigenstates of $H_\text{QM}$ and 
 $E_\lambda$ are the corresponding eigenvalues.
Introducing the density of states
\beq
\rho(E; \sigma, \vec \pi ) =
\frac{1}{V}\sum_{\lambda} \delta(E-E_\lambda (\sigma,
\vec \pi )  )
\eeq
and exploiting the fact that in all cases which are of interest to us
the eigenvalues come in pairs with opposite sign, the thermodynamic
potential becomes

\bea
\label{eq:Omega_dos}
\Omega_\text{MFA} (T,\mu; \sigma, \vec{\pi} )
=
&& - \int_0^\infty\!dE\,
\rho(E; \sigma, \vec\pi )
{\Big{\lbrace}} E + T \log\left[ 1+ e^{-\frac{E-\mu}{T}}\right] 
+ T \log\left[ 1+ e^{-\frac{E+\mu}{T}}\right]\Big\rbrace
+
\frac{1}{V}\int_V d^3x\,\mathcal{H}_\text{M}\,.
\eea

If the meson fields are homogeneous, i.e., if $\sigma(\x)$ and $\vec
\pi(\x)$ are constant in space, the quark energies can be labelled by
their conserved three-momenta $\p$,
\beq
\label{eq:Ep}
E_\p = \sqrt{ \p^2 + g^2(\sigma^2 + \vec\pi\,^2)}\,.
\eeq
 The thermodynamic potential then takes the familiar form
 \bea
\label{eq:Omegahom}
\Omega_\text{hom} (T,\mu; \sigma, \vec\pi )
=
&&
-2N_f N_c\int\!\! \frac{d^3p}{(2\pi)^3}\,
{\Big{\lbrace}} E_\p + T \log\left[ 1+ e^{-\frac{E_\p-\mu}{T}}\right] 
+ T \log\left[ 1+ e^{-\frac{E_\p+\mu}{T}}\right]\Big\rbrace
+ \frac{\lambda}{4}\left(\sigma^2 + {\vec\pi}\,^2 - v^2\right)^2,
\eea
which corresponds to the density of states
 \beq
 \label{eq:dos_hom}
 \rho_\text{hom}(E; \sigma, \vec\pi ) =  \frac{N_fN_c}{\pi^2} E
 \sqrt{E^2 - g^2(\sigma^2 + \vec\pi\,^2)}\, 
 \theta\!\left(E^2 - g^2(\sigma^2 + \vec\pi\,^2)\right)\,.
 \eeq
 For non-uniform meson fields, on the other hand, the quark
 three-momentum is not conserved and the diagonalization of
 $H_\text{QM}$ is in general very difficult.  Therefore, in this
 paper, we restrict ourselves to certain one-dimensional spatial
 modulations for which analytical solutions of the eigenvalue problem
 exist.  This will be discussed in Sec.~\ref{sec:pdinhom}.

\section{Vacuum properties}
\label{sec:vac}

In this section we discuss the vacuum properties of the model in order
to fix its parameters.  In particular, we focus on the renormalization
of the quark Dirac sea.

We assume that the vacuum is homogeneous.  For the moment, this is
simply based on phenomenology, but we will see later that the
investigated inhomogeneous solutions are indeed disfavored in vacuum.
The thermodynamic potential is then given by the $T=\mu=0$-limit of
\Eq{eq:Omegahom},
\beq
\label{eq:OmegaQMVacpisigma}
\Omega_\text{vac}(\sigma, \vec\pi) = 
- 2 N_f N_c
\int\!\! \frac{d^3p}{(2\pi)^3}\,\sqrt{\p^2 + g^2(\sigma^2 + {\vec\pi}\,^2)} 
+ \frac{\lambda}{4}\left(\sigma^2 + {\vec\pi}\,^2 - v^2\right)^2\,,
\eeq
where we have inserted the quark energies \Eq{eq:Ep}.
Minimization with respect to the sigma and pion mean fields, 
$\partial\Omega_\text{vac}/{\partial\sigma} =
\partial\Omega_\text{vac}/{\partial\pi^a} = 0$, 
and assuming that the pion fields vanish,
the nontrivial solution for the sigma field satisfies the gap equation
\beq
\label{eq:gapsigma2}
 \lambda \left(\sigma^2 - v^2\right)
=
2 N_f N_c \, g^2
\int\!\! \frac{d^3p}{(2\pi)^3}\,\frac{1}{\sqrt{\p^2 + g^2\sigma^2}} \,.
\eeq
In the following we will denote the solution of this equation by $\ave{\sigma}$.
From the Lagrangian we can then identify the constituent quark mass as
\beq
\label{eq:M}
M \equiv g\ave{\sigma}\ .  
\eeq 
For later convenience we also define the corresponding quark
propagator in Minkowski space,
 \beq 
 S_F(p) = \frac{1}{\slashed p  - M + i\epsilon} \,,
 \eeq 
 and
the loop integral
\beq
\label{eq:L1def}
L_1  = 4iN_f N_c \int\!\! \frac{d^4 p}{(2\pi)^4} \frac{1}{p^2 - M^2 +
  i\epsilon} = 2N_f N_c  \int\!\! \frac{d^3 p}{(2\pi)^3} \frac{1}{\sqrt{\p^2 + M^2}} \ ,  
\eeq 
which allows to write the gap equation as
\beq
\label{eq:gapL1}
 \lambda \left(\frac{M^2}{g^2} - v^2\right)
=
g^2\,L_1\,.
\eeq

\subsection{Standard mean-field approximation}

The thermodynamic potential as well as the momentum integral $L_1$ are
divergent and one has to specify how to treat them in order to obtain
well-defined results.  The integral is related to the Dirac sea of the
quarks and it was a standard procedure for a long time to neglect this
divergent vacuum contribution to the potential completely (``no-sea''
or ``standard mean-field'' approximation, sMFA)
\cite{Schaefer:2006ds}.  The vacuum thermodynamic potential is then
entirely given by the mesonic potential,
\beq
\label{eq:Omegansa}
\Omega_\text{vac}^\text{sMFA}(\sigma,\vec\pi) 
\equiv
U(\sigma,\vec\pi)
=
\frac{\lambda}{4}\left(\sigma^2 + {\vec\pi}\,^2 - v^2\right)^2 \,,
\eeq
and the gap equation (\ref{eq:gapL1}) simplifies to
\beq
\label{eq:gapnsa}
\frac{M^2}{g^2} - v^2 = 0\,.
\eeq

We determine the three model parameters, $g$, $\lambda$ and $v^2$ by
fitting the pion decay constant $f_\pi$, the constituent quark mass
$M$ and the sigma meson mass to fixed values. 
  $f_\pi$ is
related to $M$ and the quark-pion coupling constant $g_\pi$ by the
Goldberger-Treiman relation, \beq
\label{eq:GT}
          M = g_\pi f_\pi\,.
\eeq 
In sMFA  the coupling $g_\pi$ is equal to the bare Yukawa coupling $g$, 
which is therefore given by
\beq
\label{eq:gnsa}
g = \frac{M}{f_\pi}. 
\eeq
Inserting this into the gap equation (\ref{eq:gapnsa}) yields 
\beq
\label{eq:vnsa}
v^2 = f_\pi^2\ .
\eeq
From Eqs.~(\ref{eq:M}) and (\ref{eq:gnsa}) one also finds
$\ave{\sigma} = f_\pi$, i.e., $f_\pi$
corresponds to the vacuum expectation value of the sigma field.  As we
will see below, this does no longer hold when one goes beyond sMFA.

The meson masses in sMFA are given by the curvature of the
thermodynamic potential at the physical minimum.  Since
$\Omega_\text{vac}^\text{sMFA}$ is equal to the meson potential $U$,
see \Eq{eq:Omegansa}, this is equal to the so-called tree-level
masses, given by
\beq
\label{eq:mtree}
      m_{j,t}^2
      =     
       \left. \frac{\partial^2U}{\partial\phi_j^2}
       \right|_{\sigma = \ave{\sigma}, \vec\pi=0}\,,
\eeq 
where $\phi_j \in\{\sigma, \vec\pi\}$. 
For the sigma meson one obtains
\beq
       m_{\sigma,t}^2 
       =
       \lambda\left(3 \ave{ \sigma}^2 - v^2\right)
       =
       2\lambda f_\pi^2\ ,
\label{eq:msigmansa}
\eeq
where we employed the identities $\ave{\sigma} = v = f_\pi$ derived above.
This fixes the quartic coupling constant as 
\beq
        \lambda = \frac{\Ms^2}{2f_\pi^2}\,.
\label{eq:lambdansa}
\eeq
The pion mass vanishes,
\beq
       m_{\pi,t}^2 
       =
       \lambda\left(\ave{\sigma}^2 - v^2\right)
       =
       0\,,
\eeq
in consistency with the Goldstone theorem.

\subsection{Including the Dirac sea}

We now add the Dirac-sea contribution to the thermodynamic potential.
In this way certain quantum fluctuations which are neglected in the
sMFA are taken into account.  Since the additional term affects the
vacuum properties of the model, a refit of the parameters becomes
necessary.  We thereby assume that the divergent momentum integrals
are regularized in some way, so that the expressions can be
manipulated in a mathematically well-defined manner.  Details of the
used regularization are irrelevant at this stage and will be postponed
to Sec.~\ref{sec:parameters} where the numerical evaluations will be
discussed.

The constituent quark mass is given by the gap equation
(\ref{eq:gapL1}), now including the Dirac-sea contribution $L_1$,
which depends on $M$ as well.  We can use this equation to express the
parameter $v^2$ in terms of $M$ and the other two parameters $g$ and
$\lambda$,
 \beq v^2 =
\frac{M^2}{g^2} - \frac{g^2 L_1 }{\lambda}\,.
\label{eq:v2}
\eeq

For the determination of the meson masses and $f_\pi$ further
preparations are needed.  As in the sMFA, the tree-level masses
$m_{j,t}$ are defined via the second derivatives of the meson
potential $U$ at the minimum of the thermodynamic potential,
\Eq{eq:mtree}.  However, due to the Dirac sea, the minimum of
$\Omega_\text{vac}$ does no longer coincide with the minimum of $U$.
As a consequence, despite the same definition, the results are shifted
by $g^2\,L_1$:
\bea
       m_{\sigma,t}^2 &=& g^2\,L_1  + 2\lambda \ave{\sigma}^2\,,      
\nonumber\\
       m_{\pi,t}^2 &=& g^2\,L_1\,.
\label{eq:mt}
\eea
In particular, the tree-level pion gets massive, and we have to
include loop-corrections in order to restore the Goldstone theorem.
To this end we define the tree-level propagators in the meson channel
$j \in \left\{ \sigma, \vec\pi\right\}$
\beq
      iD_{j,t}(q^2) = \frac{i}{q^2 - m_{j,t}^2 + i\epsilon} \,, 
\eeq
and dress them with quark-antiquark polarizations loops,
\beq
       -i\Pi_j(q^2) 
       = 
       -\int\frac{d^4p}{(2\pi)^4} \mathrm{tr} \left[\Gamma_j
                            iS_F(p+q) \Gamma_j iS_F(p)\right]\,,
\label{eq:PiM}
\eeq
where $\Gamma_\sigma = \one$ and $\Gamma_{\pi^a} = i\gamma_5 \tau^a$.
Summing up a geometrical series of these self-energy insertions,
\begin{figure}[h]
\includegraphics[width=.6\textwidth]{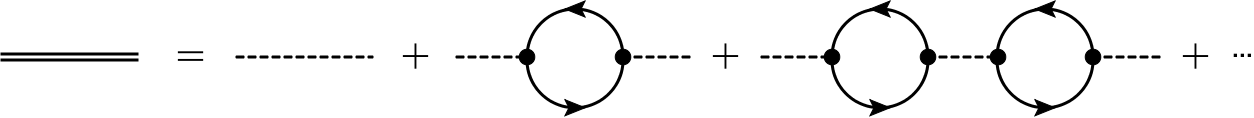}
\caption{\label{fig:polarization} Resummation of quark-antiquark
  polarization loops.  The bare meson propagators are indicated by
  dashed lines, the dressed meson propagator by the double line.  The
  filled circles indicate the bare Yukawa coupling $g$.  }
\end{figure}
as diagrammatically depicted in Fig.~\ref{fig:polarization},
we obtain the dressed meson propagator
\beq
       D_j(q^2) = \frac{1}{q^2 - m_{j,t}^2 +  g^2 \Pi_j(q^2) +i\epsilon}\ .
\label{eq:DM}
\eeq
The pole mass $m_j$ is defined by the implicit equation
\beq
       D_j^{-1}(q^2=m_{j}^2)
       \,\equalover{!}\,  0\,,
\label{mMp}
\eeq       
and we expand
\beq
       \Pi_j(q^2) = \Pi_j(m_j^2) + \Pi'(m_j^2)(q^2 - m_j^2) + \dots\,, 
\eeq
with $\Pi' \equiv \frac{d}{dq^2} \Pi(q^2)$.
The propagator in the vicinity of the pole is then given by
\beq
       D_j(q^2) = \frac{Z_j}{q^2 - m_{j}^2 +i\epsilon} \; + \; \text{regular terms}\,,
\eeq
with the wave-function renormalization constant
\beq
       Z_j = \frac{1}{1+ g^2\Pi'_j(m_j^2)}\,.
\label{eq:ZM}
\eeq

\begin{figure}[h]
\includegraphics[width=.13\textwidth]{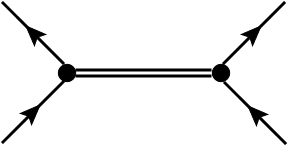}
\caption{\label{fig:meson_exc} Exchange of a dressed meson between
  quarks.}
\end{figure}

Hence, if the meson is exchanged between two quarks, as depicted in
Fig.~\ref{fig:meson_exc}, we encounter the combination
\beq 
g^2D_j(q^2) \equiv \frac{g_j^2}{q^2 - m_{j}^2+i\epsilon}\,, 
\eeq 
which allows us to identify the renormalized quark-meson coupling
constant at the pole as\footnote{ Alternatively, this result also
  follows in a straightforward way from the Lagrangian by introducing
  a wave-function renormalization for the mesons while leaving the
  quark fields unrenormalized.  } 
\beq 
g_j = g \sqrt{Z_j}\,.  
\eeq

The explicit evaluation of the polarization loops for the sigma and
pion channel is rather standard and has been presented, e.g., in the
context of mesonic excitations in the NJL model
\cite{Klevansky:1992qe}, where exactly the same integrals appear.  One
finds
\bea
\label{eq:Pipisigma}
       \Pi_\pi (q^2) &=& L_1 - \frac{1}{2}q^2 L_2(q^2)\,,
\nonumber\\
       \Pi_\sigma (q^2) 
       &=&
       L_1 -  \frac{1}{2}(q^2-4M^2) L_2(q^2)\,,
\eea
with the loop integrals $L_1$, defined in \Eq{eq:L1def}, and
\bea
\label{eq:L2def}
       L_2(q^2) 
       &=& 4iN_f N_c \int\!\! \frac{d^4 p}{(2\pi)^4} \frac{1}{[(p+q)^2
         - M^2 + i\epsilon][p^2 - M^2 + i\epsilon]} 
\nonumber\\       
       &=& 4N_f N_c \int\!\! \frac{d^3 p}{(2\pi)^3} \frac{1}{\sqrt{\p^2 + M^2}}
                \frac{1}{q^2 - 4(\p^2+M^2) + i\epsilon}\,.       
\eea
Inserting this together with the the tree-level masses, \Eq{eq:mt}, into \Eq{eq:DM}, 
we obtain the dressed meson propagators 
\bea
       D_\pi(q^2) &=& \frac{1}{q^2 \left(1-\frac{1}{2}g^2 L_2(q^2)\right) }\,,
\\
       D_\sigma(q^2) &=& 
       \frac{1}{q^2  - \frac{1}{2}g^2(q^2-4M^2)L_2(q^2) -2 \lambda M^2/g^2 }\,,       
\label{eq:Dsigmadressed}
\eea
where in the second equation we have employed \Eq{eq:M} to express the
sigma field through the constituent quark mass.  

The first equation immediately shows that the pion propagator has a pole at $q^2=0$,
i.e.,
\beq
       m_{\pi} = 0\,,
\eeq
which is consistent with the Goldstone theorem.
For the wave-function renormalization constant at the pole we read off
\beq
       Z_\pi =  \frac{1}{1-\frac{1}{2}g^2 L_2(0)}\,,
\eeq
which yields the renormalized quark-pion coupling constant  
\beq
       g_\pi^2 = \frac{g^2}{1-\frac{1}{2}g^2 L_2(0)}\,.
\eeq
Hence, employing the Goldberger-Treiman relation \Eq{eq:GT},
 the pion decay constant is given by
\beq
       f_\pi^2  
       = 
       \frac{M^2}{g^2}\left(1-\frac{1}{2}g^2 L_2(0)\right)\,,
\label{eq:fpi}
\eeq
which can be used to fix the Yukawa coupling $g$:
\beq
       g^2 = \frac{M^2}{f_\pi^2 + \frac{1}{2}M^2 L_2(0)}\,.
\label{eq:g2}
\eeq

\noindent
For the pole mass of the sigma meson, \Eq{eq:Dsigmadressed} yields the relation
\beq
       m_{\sigma}^2 
       -\frac{1}{2} g^2 (m_{\sigma}^2 -4M^2) L_2(m_{\sigma}^2)
       -2\lambda \frac{M^2}{g^2}       
       = 0\,,
\label{eq:msigma}
\eeq
which can be used to fix the quartic coupling constant: 
\beq
      \lambda
       =
       2g^2 \frac{m_{\sigma}^2}{4M^2} 
       \left[ 1
       - \frac{1}{2}g^2
       \left(1 - \frac{4M^2}{m_{\sigma}^2} \right) L_2(m_{\sigma}^2)\right]\,.
\label{eq:lambda}
\eeq
In the NJL model the sigma-meson mass in the chiral limit is always
equal to $2M$~\cite{Klevansky:1992qe}.  In the QM model there is no
such restriction, but for $m_\sigma = 2M$ the expression for $\lambda$
becomes particularly simple: The $L_2$-term drops out, and we find
$\lambda = 2g^2$.

For arbitrary sigma masses it is useful to separate
\beq
       L_2(m_\sigma^2) = L_2(0) + \delta L_2(m_\sigma^2)\,,
\label{eq:L2split}
\eeq
where $\delta L_2(m_\sigma^2)$ is a finite expression.
We can then employ \Eq{eq:fpi} to express $L_2(0)$  through $f_\pi$
and get
\beq
      \lambda
       =
       2g^2 \left[  1  +  g^2 \left(\frac{m_\sigma^2}{4M^2} -1\right)
       \left(\frac{f_\pi^2}{M^2} - \frac{1}{2}\delta L_2(m_{\sigma}^2)
       \right)
       \right]\,.
\label{eq:lambdav2}
\eeq

Finally, we briefly comment on the so-called screening masses of the
mesons which are defined in analogy to the tree-level masses,
\Eq{eq:mtree}, but with the meson potential $U$ replaced by the
thermodynamic potential $\Omega$, i.e., in vacuum,
\beq
\label{eq:mscreen}
      m_{j,s}^2
      =     
       \left. \frac{\partial^2\Omega_\text{vac}}{\partial\phi_j^2}
       \right|_{\sigma = \ave{\sigma}, \vec\pi=0}\,.
\eeq 
These masses are related to the unrenormalized dressed meson propagator for vanishing
momentum, $D_j(0) = -1/m_{j,s}^2$, so that $m_{\pi,s} = 0$, in agreement with the pole mass.
For the sigma meson, on the other hand, one finds 
\beq
\label{eq:mss}
       m_{\sigma,s}^2 = 2\lambda \frac{M^2}{g^2} - 4M^2 + 4g^2f_\pi^2
       \,= \frac{g^2f_\pi^2}{M^2} m_\sigma^2 + (4M^2-m_\sigma^2)\frac{1}{2}g^2 \delta L_2(m_\sigma^2)
       \ ,
\eeq
which coincides with the pole mass in the sMFA, but is different when
the quark loops are taken into account.  Hence, fixing the screening
mass of the sigma rather than the pole mass, as was done, e.g., in
Ref.~,\cite{Skokov:2010sf}, leads in general to a different value for
$\lambda$.  However, we consider the pole mass as the more physical
quantity (see also Ref.~\cite{Strodthoff:2011tz}) and fix $\lambda$
from \Eq{eq:lambdav2}.

In summary, the model parameters $v^2$, $g$, and $\lambda$ are
determined by the equations (\ref{eq:v2}), (\ref{eq:g2}), and
(\ref{eq:lambdav2}), respectively, for fixed values of $M$, $m_\sigma$
and $f_\pi$, and a given regularization procedure to tame the loop
integrals $L_1$ and $L_2$. This program will be discussed next in detail.

\subsection{Regularization and parameter fixing}
\label{sec:parameters}

For the regularization of the quark loops, we employ the Pauli-Villars
(PV)-inspired scheme developed in Ref.~\cite{Nickel:2009wj}.  In this
scheme, the divergent term in the general expression for the
mean-field thermodynamic potential, \Eq{eq:Omega_dos}, is replaced in
the following way:
\beq
\label{eq:PV_rho}
- \int_0^\infty\!dE\, \rho(E; \sigma, \vec\pi ) \, E 
\,\rightarrow\, 
- \int_0^\infty\!dE\, \rho(E; \sigma, \vec\pi ) \, \sum_{j=0}^3 c_j \sqrt{E^2 + j\Lambda^2} \,, 
\eeq
with a cutoff parameter $\Lambda$ and the coefficients $c_0 = 1$, $c_1 = -3$, $c_2 = 3$,
and $c_3 = -1$.
In homogeneous matter,
following the same steps as before, this amounts to the standard PV regularization, 
\beq
       \int\!\! \frac{d^3 p}{(2\pi)^3} \,F(M^2)
       \,\rightarrow\,        
       \int\!\! \frac{d^3 p}{(2\pi)^3} \sum_{j=0}^3 c_j \, F(M^2 + j\Lambda^2)
\eeq
of the loop integrals $L_1$ and $L_2(q^2)$. 
Explicitly, one finds
\bea
L_1 &\rightarrow&
\frac{N_f N_c}{4\pi^2} \sum_{j=0}^3 c_j M^2_j \ln
M^2_j\,,
\\[1mm] 
L_2(0)  &\rightarrow&
\frac{N_f N_c}{4\pi^2} \sum_{j=0}^3 c_j \ln M^2 _j\,,
\\[1mm]
\mathrm{Re}\,\delta L_2(q^2)  &\rightarrow&
\frac{N_f N_c}{4 \pi^2}
\sum_{j=0}^3 c_j \, \left\{ f\left(\frac{4 M^2_ j}{q^2} -1\right)
-2 \right\},
\\[1mm]
\mathrm{Im}\,\delta L_2(q^2) &\rightarrow&
-\frac{N_f N_c}{4 \pi}
\sum_{j=0}^3 c_j \, \sqrt{1 -\frac{4 M^2_ j}{q^2}} \,
\theta(q^2-4M^2)\,, 
\eea 
with $M_j^2 = M^2 + j\Lambda^2$ and
\beq f(x) = \left\{ \begin{array}{ll}
    2\sqrt{x} \arctan(\frac{1}{\sqrt{x}})\,, & \quad x > 0
    \\[1mm]
    \sqrt{-x} \ln(\frac{1+\sqrt{-x}}{1-\sqrt{-x}})\,, & \quad x < 0
    \\[1mm]
    0\,, & \quad x=0\,.
  \end{array} \right.                 
\eeq
Similar to \Eq{eq:L2split}, we have split the function $L_2(q^2)$ into
two parts, $L_2(0)$ and $\delta L_2(q^2)$.  Since the latter is
finite, its regularization is not necessary.  We prefer, however, to
regularize the entire function $L_2(q^2)$.  This has the advantage
that the sMFA results are recovered when the limit $\Lambda
\rightarrow 0$ is taken.  For $\Lambda \rightarrow \infty$, on the
other hand, the regularization of $\delta L_2(q^2)$ makes no
difference, as the regulator terms, corresponding to $j = 1,2,3$,
vanish, and only the $j=0$ term survives.

We also note that $\delta L_2$ receives an imaginary part above the
two-quark threshold, $q^2 > 4M^2$.  This becomes relevant for
$m_\sigma > 2M$ and signals a non-vanishing width of the sigma meson
due to quark-antiquark decay.  In the parameter fit we will then
define the sigma mass as the point where the real part of the inverse
sigma propagator vanishes.

As an alternative to the PV prescription, we could regularize the
divergencies with a sharp three-momentum cutoff.  For homogeneous
phases both schemes give similar results, as will be shown in
Sec.~\ref{sec:pdhom}.  We note, however, that a momentum cutoff is not
appropriate for inhomogeneous phases.  Intuitively this can be seen
from the fact that a restriction of the individual quark momenta would
also restrict the possible wave numbers of the inhomogeneity.  A more
formal argument will be given in Sec.~\ref{sec:GL} in the context of a
Ginzburg-Landau analysis.

Throughout this paper, we fix the vacuum values of the constituent
quark mass to $M = 300$~MeV and of the pion decay constant to its
chiral-limit value $f_\pi = 88$~MeV.  For the sigma-meson mass we will
take the NJL-model relation $m_\sigma = 2M = 600$~MeV .  Later,
in Sec.~\ref{sec:sigmamasses}, $m_\sigma$ will be varied in order to
study its influence on the phase diagram.

\begin{figure}[h]
\centering
\includegraphics[width=.325\textwidth]{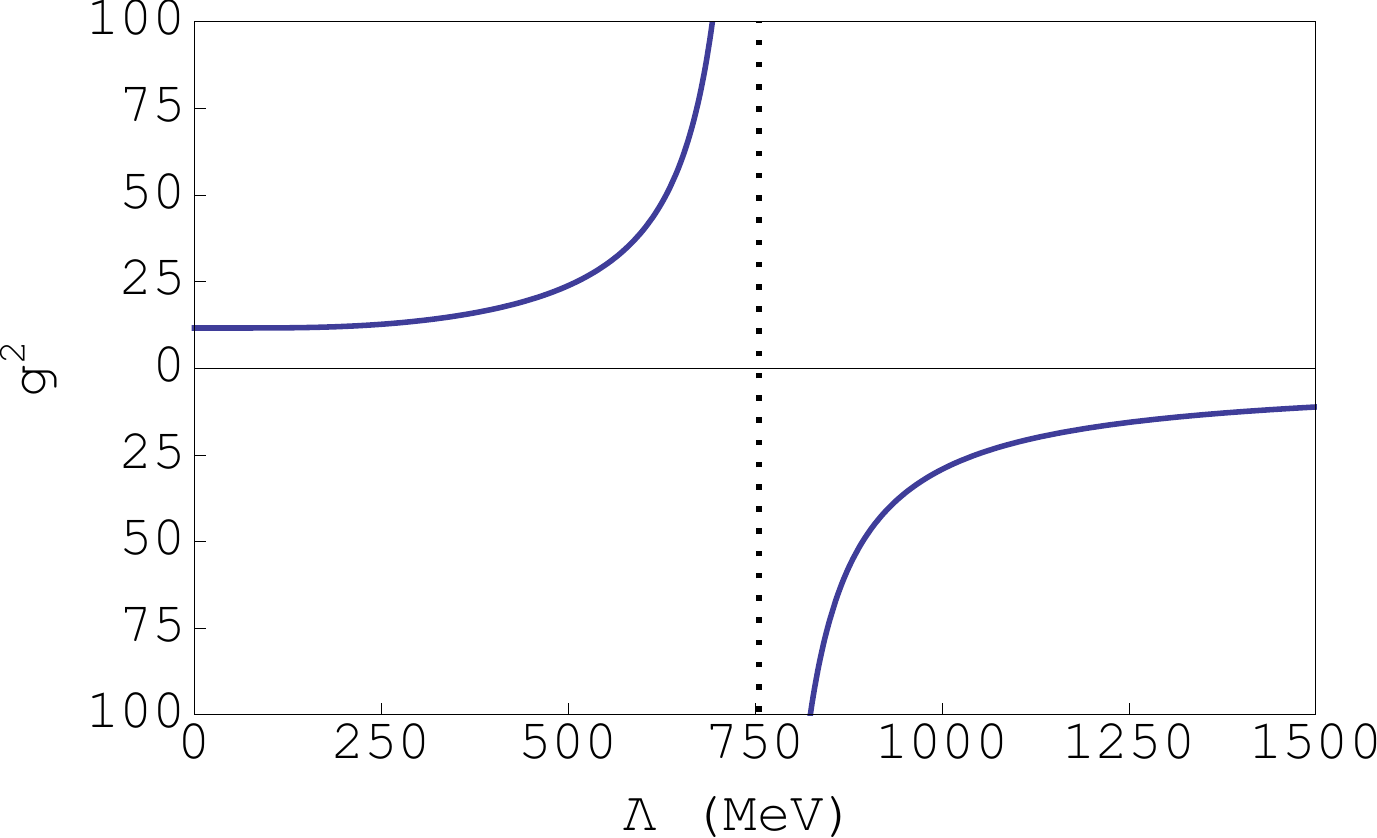}
\includegraphics[width=.325\textwidth]{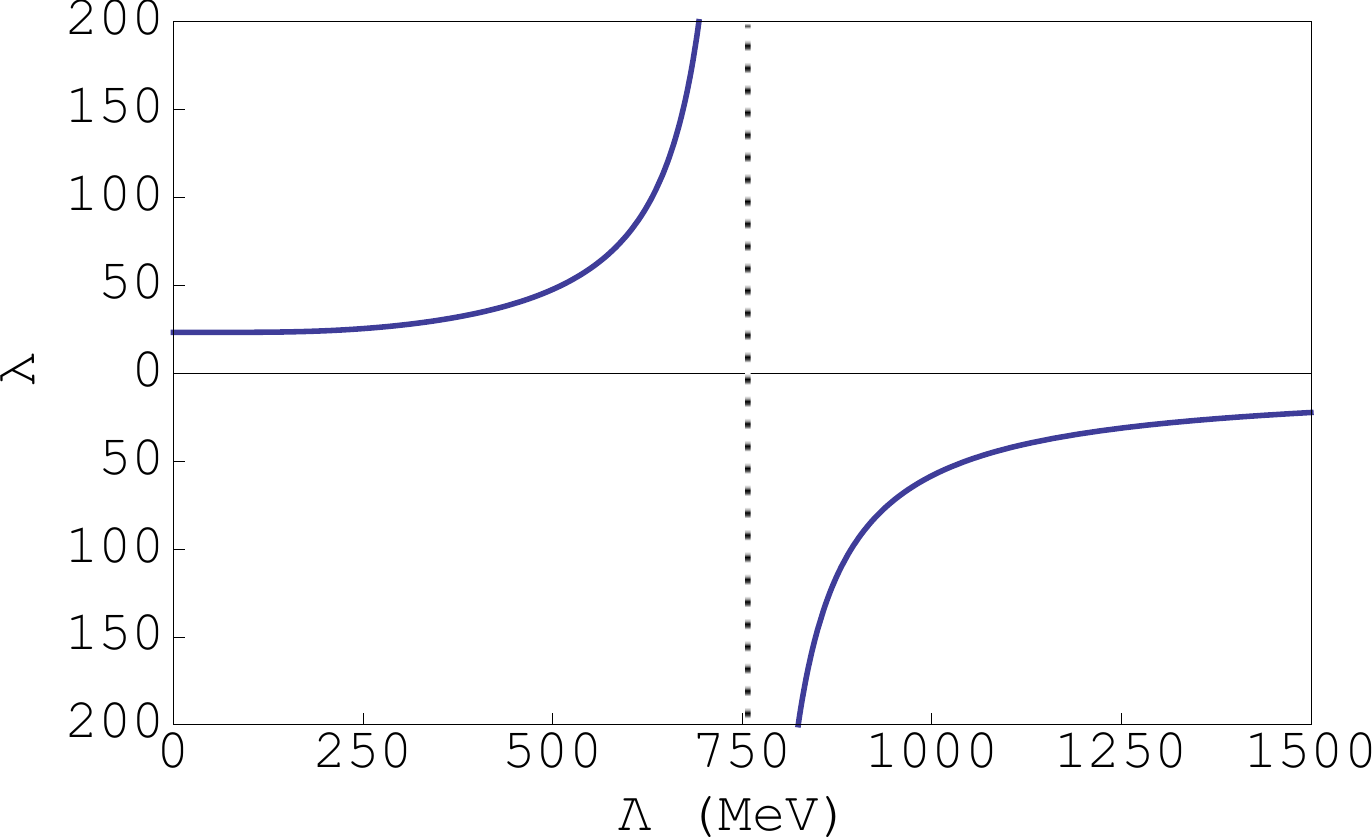}
\includegraphics[width=.325\textwidth]{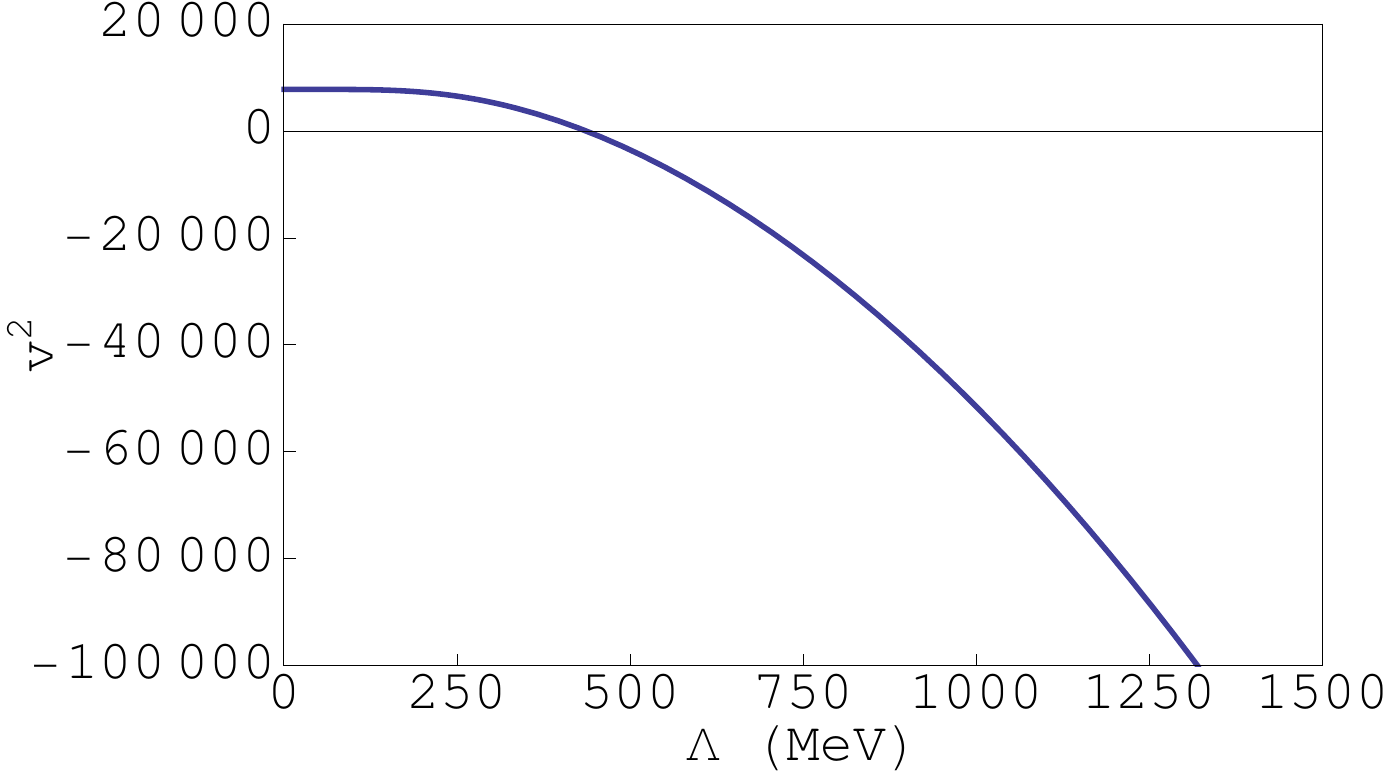}
\caption{ The model parameters $g^2$, $\lambda$ and $v^2$ (left to
  right) as functions of the PV cutoff parameter.  }
\label{fig:qmparams}
\end{figure}

Numerical results for the parameters as functions of the cutoff in the
PV regularization scheme are shown in Fig.~\ref{fig:qmparams}.  For a
sharp $O(3)$ cutoff the behavior is qualitatively similar.  At a
certain value of $\Lambda$, the couplings $g^2$ and $\lambda$ diverge
and change their sign.  
As can be seen from Eqs.~(\ref{eq:g2}) and (\ref{eq:lambda}),
  this happens at the point where
$f_\pi^2 = -\frac{1}{2}M^2 L_2(0)$. Interestingly, this is exactly the expression for 
$f_\pi^2$ in the NJL model~\cite{Klevansky:1992qe}.
The parameter $v^2$ changes its sign as well, although at a different
value of $\Lambda$.  Above this point the chiral-symmetry breaking is
no longer caused by the meson potential, like in the sMFA, but by the
Dirac sea of the quarks.  As one can see from Eqs.~(\ref{eq:g2}) and
(\ref{eq:lambdav2}) the coupling constants $g^2$ and $\lambda$
approach zero when the cutoff goes to infinity.  It follows from
\Eq{eq:mss} that the screening mass of the sigma meson vanishes as
well in this limit, whereas the pole mass stays constant by
construction.
Moreover,  since $\mathrm{Im}\,\delta L_2(q^2)$ stays finite 
while $g^2$ goes to zero,
the decay width of the sigma meson into two quarks also vanishes,
even for $m_\sigma > 2M$.

\section{Numerical results for homogeneous phases}
\label{sec:pdhom}

We now extend the calculations to nonvanishing temperature and
chemical potential.  As a first step, and in order to set the stage
for our subsequent studies of inhomogeneous phases, we restrict
ourselves to homogeneous order parameters.  To this end, the
thermodynamic potential, \Eq{eq:Omegahom}, is minimized, now including
the $T$- and $\mu$-dependent terms.  Since these terms are convergent,
we leave them unregularized (as customary in the sMFA) and only
regularize the vacuum term in the way discussed in the previous
section.

\begin{figure}[h]
\centering
\includegraphics[angle=270,width=.7\textwidth]{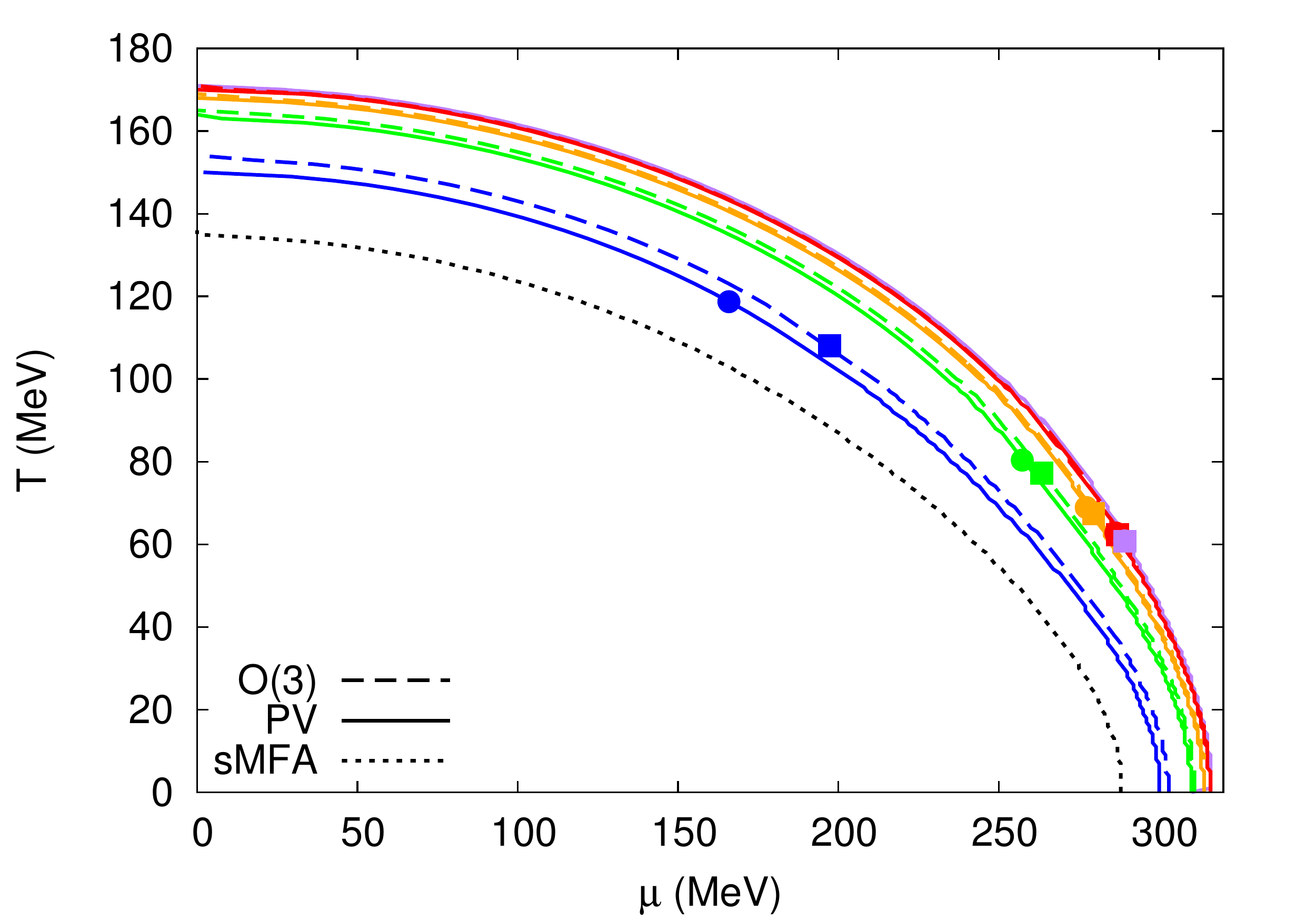}
\caption{ Phase diagrams for homogeneous condensates.  The dotted line
  indicates the first-order phase boundary in the sMFA. For the other
  lines the Dirac sea was included, using PV regularization (solid
  lines) or a sharp three-momentum cutoff (dashed lines), with
  different values of the cutoff: $\Lambda = 300$ MeV, 600~MeV, 1~GeV,
  2~GeV, 5~GeV (from bottom to top).  The filled circles (for PV) and
  squares (for three-momentum cutoff) indicate the tricritical points,
  separating second-order (to the left) from first-order (to the
  right) phase transitions.  }
\label{fig:allpdhom}
\end{figure}

Numerical results for the phase diagram are presented in
\Fig{fig:allpdhom}.  The various solid and dashed lines indicate the
phase boundaries obtained with Pauli-Villars regularization and
three-momentum cutoff, respectively, using different values of
$\Lambda$.  The dotted line corresponds to the sMFA result.  It is
well known that in the chiral limit and without the Dirac sea
contribution, the Quark-Meson model exhibits a first-order transition
along the entire phase boundary, i.e., even at vanishing chemical
potential \cite{Schaefer:2006ds, Schaefer:2004en}.  This problem was
carefully analyzed in Ref.~\cite{Skokov:2010sf}, where it was
identified as an artifact of the sMFA, which is cured when the
Dirac-sea contributions are included.  Indeed, for both regularization
schemes we find that already for $\Lambda =$ 300~MeV there is a
tricritical point (CP), which separates the first-order transition at
higher $\mu$ from a second-order one at low $\mu$, both for PV
regularization and for a three-momentum cutoff.  When $\Lambda$ is
increased further, the second-order part of the transition line
continues to grow, so that the CP moves to higher chemical potentials.
This can be interpreted as the effect of the vacuum fluctuations,
which smoothen the transition \cite{Skokov:2010sf}.  The fluctuations
also increase the ``bag pressure'', i.e., the pressure difference
between the chirally broken and restored vacuum solutions.  As a
consequence, the phase boundary is shifted to higher temperatures and
chemical potentials.

At the same time, the phase diagrams for the two different
regularizations become more and more similar, until beyond $\Lambda
\approx 1$ GeV the transition lines and the location of the critical
point basically overlap. This is of course the expected behavior for a
renormalizable model: with a proper rescaling of the parameters, at
high values of $\Lambda$ the results become cutoff-independent. For
practical purposes, in the following we will interpret results
calculated at a $\Lambda = 5$ GeV cutoff to be the ``renormalized''
ones. A numerical cross-check performed at $\Lambda = 10$ GeV confirms
the stability of these results.

\section{Ginzburg-Landau analysis}
\label{sec:GL}

Before extending our numerical calculations to investigate
inhomogeneous phases, we first want to explore the possible phase
structure within a Ginzburg-Landau (GL) approach.  In particular we
are interested in the Lifshitz point (LP), i.e., the point where the
inhomogeneous phase and the two homogeneous phases with broken and
restored chiral symmetry meet, and its location relative to the
tricritical point (CP), where the second-order chiral phase transition
turns into first order if the analysis
 is restricted to homogeneous phases.  Within the GL
approach, this can be analyzed in a rather general way, without
specifying the explicit form of the inhomogeneity~\cite{Nickel:2009ke,
  Nickel:2009wj, Abuki:2011pf}.

In the following, we basically extend the analysis of
Ref.~\cite{Nickel:2009wj} to include the Dirac sea.  For this, the
mean-field thermodynamic potential, \Eq{eq:Omega1QM}, is expanded
around the symmetric ground state in terms of the order parameters and
their gradients.  We restrict ourselves to a single isospin component
of the pion field, $\pi^a(\x) =\pi(\x)\delta_{a3}$, which is
sufficient to describe the phases we are interested in.  The
thermodynamic potential can then be expressed in terms of a complex
constituent-quark mass function
\beq
       M(\x) = g\left(\sigma(\x) + i\pi(\x)\right)\,,
\eeq 
and the GL expansion takes the form
\beq
       \Omega_\text{MFA}(T,\mu;M) = \Omega_\text{MFA}(T,\mu;0)
       + \frac{1}{V}\int d^3x\, \left\{\frac{1}{2}\gamma_2 |M(\x)|^2
       + \frac{1}{4}\gamma_{4,a} |M(\x)|^4 + \frac{1}{4}\gamma_{4,b} |\nabla M(\x)|^2
       + \dots
       \right\}\,.
\eeq
As in Refs.~\cite{Nickel:2009ke,
  Nickel:2009wj}, we assume that terms of order 6 or higher are positive. As we will see later, the 
predictions based on this assumption are confirmed by direct numerical computations of the phase diagram.

In order to compute the GL coefficients $\gamma_i$ we decompose
$\Omega_\text{MFA}$ into  a quark-loop contribution given by
\beq
        \Omega_{q\bar q}(T,\mu;M) = - \frac{T}{V} \mathbf{Tr} \, 
                              \mathrm{Log}\left(\frac{1}{T}\left(i\partial_0 +H_\text{QM}-\mu\right)\right) \,,
\eeq
and a pure meson contribution
\beq 
       \Omega_\text{M}(M) = \frac{1}{V}\int_V\!\! d^3x\,\mathcal{H}_\text{M}\,, 
\eeq
and expand both parts in the same manner,
\bea
      \Omega_{q\bar q} (T,\mu; M) &=& \Omega_{q\bar q}(T,\mu; 0) 
       + \frac{1}{V}\int_V\!\! d^3x\, \left\{\frac{1}{2}\beta_2 |M(\x)|^2
       + \frac{1}{4}\beta_{4,a} |M(\x)|^4 + \frac{1}{4}\beta_{4,b} |\nabla M(\x)|^2
       + \dots
       \right\}\,,
\\       
       \Omega_\text{M}(M) &=& \Omega_\text{M}(0) 
       + \frac{1}{V}\int_V\!\! d^3x\, \left\{\frac{1}{2}\alpha_2 |M(\x)|^2
       + \frac{1}{4}  \alpha_{4,a} |M(\x)|^4 + \frac{1}{4} \alpha_{4,b} |\nabla M(\x)|^2
       + \dots \right\}\,,      
\eea
so that $\gamma_i = \alpha_i + \beta_i$.

The mesonic coefficients are easily obtained from Eqs.~(\ref{eq:Uqm}) and (\ref{eq:Hmes}), 
yielding 
\beq
       \alpha_2 = -\frac{\lambda v^2}{g^2}\,, \quad
       \alpha_{4,a} = \frac{\lambda}{g^4}\,, \quad
       \alpha_{4,b} = \frac{2}{g^2}\,, 
\eeq
while, except for a trivial constant in $\beta_2$, the quark-loop
terms are the same as in the NJL model~\cite{Nickel:2009wj, Nickel:2009ke}.  
They are given by
\bea
\beta_2 &=& \beta_2^\text{vac} + \beta_2^\text{med}\,,
\\
\beta_{4,a} = \beta_{4,b} &=& \beta_4^\text{vac} + \beta_4^\text{med}\,,
\eea
with the vacuum parts
\bea
        \beta_2^\text{vac} &=& \big.-L_1\big|_{M=0}\ ,
\\[1ex]
        \beta_4^\text{vac} &=& \big.-L_2(0)\big|_{M=0}\ , 
\eea
and the explicitly $T$- and $\mu$-dependent medium parts
\bea
        \beta_2^\text{med} &=& 2N_f N_c \int\!\! \frac{d^3p}{(2\pi)^3}\, \frac{1}{|\p|} 
        \left(n(\p) + \bar{n}(\p)\right)\,,
\\
        \beta_4^\text{med} &=& -N_f N_c \int\!\! \frac{d^3p}{(2\pi)^3}\, \frac{1}{|\p|^3} 
        \left(n(\p) + \bar{n}(\p) + \frac{|\p|}{T} \Bigl[ n(\p)
            \left(1-n(\p)\right) + \bar n(\p) \left(1-\bar n(\p) \right)\Bigr]
        \right)\,,
\eea
where $n(\p) = 1/[e^{(|\p| -\mu)/T}+1]$ and $\bar n(\p) = 1/[e^{(|\p| +\mu)/T}+1]$
are Fermi distribution functions for massless quarks and antiquarks, respectively.

The coefficients $\beta_4^\text{vac}$ and $\beta_4^\text{med}$ both
exhibit logarithmic divergencies in the infrared, which cancel each
other if they are summed.  In the sMFA, where the vacuum term is
absent, this cancellation does not occur and this originates the
artifact that the first-order phase transition extends up to $\mu =
0$~\cite{Skokov:2010sf}.  In addition, the vacuum parts of the $\beta$
coefficients are ultraviolet divergent, as obvious from the presence
of the loop integrals $L_1$ and $L_2$.  In order to be consistent with
our numerical studies of the phase diagram, these integrals are
regularized in the same way as described in Sec.~\ref{sec:parameters},
while the UV-finite medium parts are left unmodified.

The most important result is that $\beta_{4,a}$ and $\beta_{4,b}$ are
equal\footnote{The proof of this equality requires an integration by
  parts and relies on the assumption that there are no surface
  contributions to the integral. This assumption would be violated by
  introducing a sharp momentum cutoff, which is our main reason not to
  use cutoff regularization for inhomogeneous phases.}, with the
consequence that in the NJL model the LP coincides with the CP
\cite{Nickel:2009ke}.  Because of the $\alpha$-coefficients, this is
not necessarily true in the QM model.  In general, the CP is given by
the condition that the coefficients of the quadratic and quartic terms
vanish, while at the LP the quadratic and the gradient terms are zero,
\bea
\big. \gamma_2 \big|_\text{CP} &=& \big. \gamma_{4,a}
\big|_\text{CP} = 0\,,
\label{eq:GLCP}
\\[1mm]
\big. \gamma_2 \big|_\text{LP} &=& \big. \gamma_{4,b}
\big|_\text{LP} = 0\,.  
\label{eq:GLLP}
\eea 
Since $\beta_{4,a}=\beta_{4,b}\equiv\beta_4$, both conditions are fulfilled
simultaneously if $\alpha_{4,a}=\alpha_{4,b}$, i.e., if $\lambda =
2g^2$.  As can be seen from \Eq{eq:lambda}, this is the case if
$m_\sigma = 2M$, which is just the NJL-model relation between sigma
and constituent quark mass in the chiral limit.  This result is quite
remarkable. It means that if $m_\sigma$ and $M$ fulfill the NJL-model
relation, the LP and the CP coincide, just as they do in the NJL
model.\footnote{For the sMFA, this result is basically contained in the
  expressions derived in Ref.~\cite{Nickel:2009wj}, but was missed due
  to a simple algebraic mistake.}
We note that we would not have obtained this result if we had determined $\lambda$
by fitting the screening mass of the sigma meson instead of the pole mass.

In contrast to the NJL model, however, the ratio of $m_\sigma$ and $M$ can be chosen freely
in the QM model, and we thus have the possibility to 
separate the two points. 
At the LP, the coefficient $\gamma_{4,a}$, which vanishes at the CP, takes the value
\beq
       \big.\gamma_{4,a}\big|_\text{LP} =  \alpha_{4,a}-\alpha_{4,b} 
=    
       2 \left( \frac{m_{\sigma}^2}{4M^2}-1 \right)
       \left( \frac{f_\pi^2}{M^2} - \frac{1}{2} \delta L_2(m_{\sigma}^2) \right)\,,
\eeq
where we have used \Eq{eq:lambdav2} to eliminate the coupling constants.
For $\Ms$ close to $2M$ we can then expand
\beq
       0 = \left.\gamma_{4,a}\right|_\text{CP} =
             \left.\gamma_{4,a}\right|_\text{LP}  
             + \left.\frac{\partial\beta_4}{\partial T}\right|_\text{LP} \Delta T
             + \left.\frac{\partial\beta_4}{\partial \mu}\right|_\text{LP} \Delta\mu \,,
\eeq
where $\Delta T = T_\text{CP}-T_\text{LP}$ and $\Delta \mu = \mu_\text{CP}-\mu_\text{LP}$ are the 
temperature and chemical-potential shifts of the CP relative to the LP.
At the same time we must require that $\gamma_2$ stays zero, i.e.,
\beq
        \left.\frac{\partial\beta_2}{\partial T}\right|_\text{LP} \Delta T
        + \left.\frac{\partial\beta_2}{\partial \mu}\right|_\text{LP} \Delta\mu = 0\,,
\eeq
so that we obtain for the shifts:
\bea
       \Delta T &=& \left.\left( \frac{\frac{\partial\beta_2}{\partial\mu}}
                         {    \frac{\partial\beta_2}{\partial T} \frac{\partial\beta_4}{\partial\mu}
                           -   \frac{\partial\beta_2}{\partial\mu}\frac{\partial\beta_4}{\partial T}  }\,
                         \gamma_{4,a}\right)\right|_\text{LP}  \,,
\\    
       \Delta \mu &=& \left.\left( \frac{-\frac{\partial\beta_2}{\partial T}}
                         {    \frac{\partial\beta_2}{\partial T} \frac{\partial\beta_4}{\partial\mu}
                           -   \frac{\partial\beta_2}{\partial\mu}\frac{\partial\beta_4}{\partial T}  }\,
                         \gamma_{4,a}\right)\right|_\text{LP}  \,.                      
\eea
Hence, even without working out the $T$ and $\mu$ derivatives of the
$\beta$-coefficients, it becomes clear that the sign change of
$\gamma_{4,a}|_\text{LP}$ at $\Ms = 2M$ implies that $\Delta T$ and
$\Delta \mu$ change their signs as well. This strongly suggests that
we can realize two different scenarios: one where the CP lies inside
the inhomogeneous phase and one where it is outside.  We will
investigate this numerically in Sect.~\ref{sec:sigmamasses}.  Until
then, we will concentrate on the NJL-like case $m_\sigma = 2M$.

\section{Inhomogeneous phases}
\label{sec:pdinhom}

We now proceed with the numerical study of the phase diagram,
including the possibility of inhomogeneous phases, i.e., allowing the
mesonic mean fields to be non-uniform in space.  For simplicity, we
will restrict ourselves to certain one-dimensional modulations for
which analytical expressions for the quark spectral density
$\rho(E;\sigma,\vec\pi)$ are known.  In particular, we will focus on
the chiral density wave (CDW) ansatz, which amounts to a
one-dimensional plane-wave type modulation of the chiral condensate
\cite{Sadzikowski:2000ap, Nakano:2004cd, Nickel:2009wj}.  Choosing, as
customary, the modulation to point to the $z$ direction, the mesonic
mean fields are given by

\beq
\sigma(z) = \frac{\Delta}{g} \cos(2qz) \,, \qquad \pi(z) = \frac{\Delta}{g} \sin(2qz) \,,
\label{eq:sigma_pi_CDW}
\eeq
where the pion field is again restricted to the third isospin component,
$\pi^a(z) = \pi(z)\delta_{a3}$. 
The mean fields can be combined into the complex order parameter
\beq
M(z) =  g(\sigma(z) + i\pi(z)) = \Delta e^{2iqz}\,.
\label{eq:Mcdw}
\eeq
Hence, the modulation is characterized by two variational parameters,
an amplitude $\Delta$ and a wave number $q$, which can both be chosen
to be real and nonnegative.  The factor of 2 in the exponent is just
to be consistent with current conventions (e.g., \cite{Basar:2009fg,
  Nickel:2009wj}).  The corresponding quark spectral density reads
\cite{Nickel:2009wj}
\bea
\label{eq:rho_CDW}
\rho(E;\sigma,\vec\pi) = {\rho}_\text{CDW}(E;\Delta,q)
=
N_f N_c \,\frac{E}{2\pi^2} \Big\{&&
\theta(E-q-\Delta) \sqrt{(E-q)^2-\Delta ^2} +
\nonumber\\
&&
\theta(E-q+\Delta)\,\theta(E+q-\Delta)\sqrt{(E+q)^2-\Delta ^2}+
\nonumber\\
&&
\theta(q-\Delta-E)
\left(\sqrt{(E+q)^2-\Delta^2}-\sqrt{(E-q)^2-\Delta^2}\right)
\Big\}
\,,
\eea
while for the meson contribution to the thermodynamic potential we have
\bea
\label{eq:Omega_M_CDW}
\frac{1}{V}\int d^3x\, \mathcal{H}_\text{M} 
&=&\frac{1}{L} \int_0^L dz
\left\lbrace-\frac{1}{2}\left[-\left(\frac{\partial\sigma}{\partial
        z}\right)^2 -\left(\frac{\partial\pi}{\partial
        z}\right)^2\right] + \frac{\lambda}{4} \left( \sigma^2 + \pi^2
    - v^2 \right)^{2} \right\rbrace  
\nonumber\\
&=& 
 \frac{1}{2} \left(\frac{2q \Delta}{g}\right)^2 + \frac{\lambda}{4}
 \left[ \left(\frac{\Delta}{g}\right)^2 - v^2\right]^2 \,, 
 \label{eq:mesonpotCDW}
\eea
where $L = \pi/q $ is the period of the spatial modulation.
Inserting this into \Eq{eq:Omega_dos}, we obtain the thermodynamic potential
\beq
       \Omega_\text{CDW}(T,\mu;\Delta,q) \equiv 
       \Omega_\text{MFA}(T,\mu;\sigma,\vec\pi)\,
\eeq 
with $\sigma$ and $\vec\pi$ given by \Eq{eq:sigma_pi_CDW}.  As in the
homogeneous case, which is contained in the above expressions as the
limit $q\rightarrow 0$, we only regularize the vacuum part, now
exclusively employing the PV scheme, \Eq{eq:PV_rho}.

It can be shown on general grounds that for small values of the wave
number $q$ the thermodynamic potential at zero temperature and
chemical potential rises proportional to $f_\pi^2 q^2$, when expanded
around the homogeneous ground state \cite{Broniowski:1990gb}.  More
precisely, when we define
\beq
      \Delta\Omega_\text{vac}(q)
      \equiv
      \Omega_\text{CDW}(0,0;M,q) -  \Omega_\text{CDW}(0,0;M,0)\,,  
\eeq
with $M$ being the constituent quark mass in vacuum,
we expect that
\beq
\Delta\Omega_\text{vac}(q)
=
\frac{1}{2}f_\pi^2 (2q)^2 + {\mathcal O}(q^4)
\,.
\label{eq:deltaomega}
\eeq
In order to show this, we split $\Delta\Omega_\text{vac}$ into the
Dirac-sea contribution of the quarks and the kinetic contribution of
the mesons,
\beq
\Delta\Omega_\text{vac}(q) =
\Delta\Omega_\text{vac,$q \bar q  $}(q)
+
\Delta\Omega_\text{vac,M}(q) 
\,,
\eeq
with
\beq
\Delta\Omega_\text{vac,$q\bar q$}(q) = 
- \int_0^\infty \!\! dE
\Big(\rho_\text{CDW}(E; M, q) -\rho_\text{CDW}(E; M,0)\Big) 
\sum_j c_j \sqrt{E^2 + j\Lambda^2} 
\label{eq:deltaomegasea}
\eeq
and 
\beq
\Delta\Omega_\text{vac,M}(q) =
\frac{1}{2} \left(\frac{2q M}{g}\right)^2 \ .
\eeq
In the sMFA, we have $\Delta\Omega_\text{vac,$q \bar q$} = 0$, and
\Eq{eq:deltaomega} follows immediately from the Goldberger-Treiman
relation, \Eq{eq:gnsa}.  When the Dirac sea is included, one finds
\beq 
\Delta\Omega_\text{vac,$q \bar q$}(q) = -M^2 L_2(0) q^2 + {\mathcal
  O}(q^4) \,,
\label{eq:deltaomegaq}
\eeq
(see appendix) and  \Eq{eq:deltaomega} is obtained with \Eq{eq:fpi}.
This demonstrates that our treatment of inhomogeneous 
phases is consistent with the parameter fixing in vacuum.

\begin{figure}[htp]
\centering
\includegraphics[width=.4\textwidth]{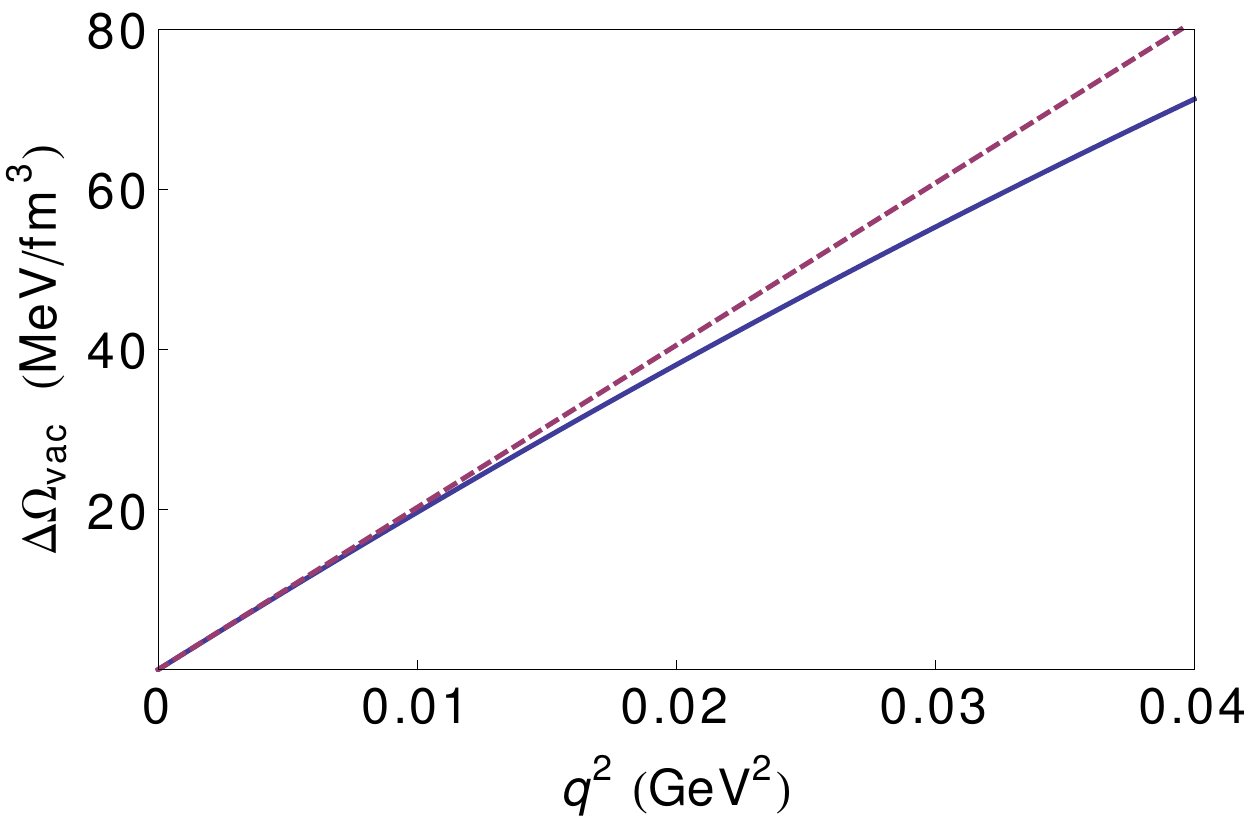}
\caption{$\Delta\Omega_\text{vac}$ as a function of $q^2$ (solid line), 
compared with the leading-order behavior $\frac{1}{2}f_\pi^2 (2q)^2$ (dashed line).
The calculations have been performed with $\Lambda=5$ GeV. 
}
\label{fig:cfrQ2}
\end{figure}

Our numerical results for $\Delta\Omega_\text{vac}$ are displayed in
Fig.~\ref{fig:cfrQ2} against $q^2$. At low momenta the agreement with
$\frac{1}{2} f_\pi^2 (2q)^2$ (dashed line) is excellent, while at
higher wave numbers the higher-order contributions become visible.

\subsection{Phase diagram for $m_\sigma = 2M$}

\begin{figure}[htp]
\centering
\includegraphics[angle=270,width=.32\textwidth]{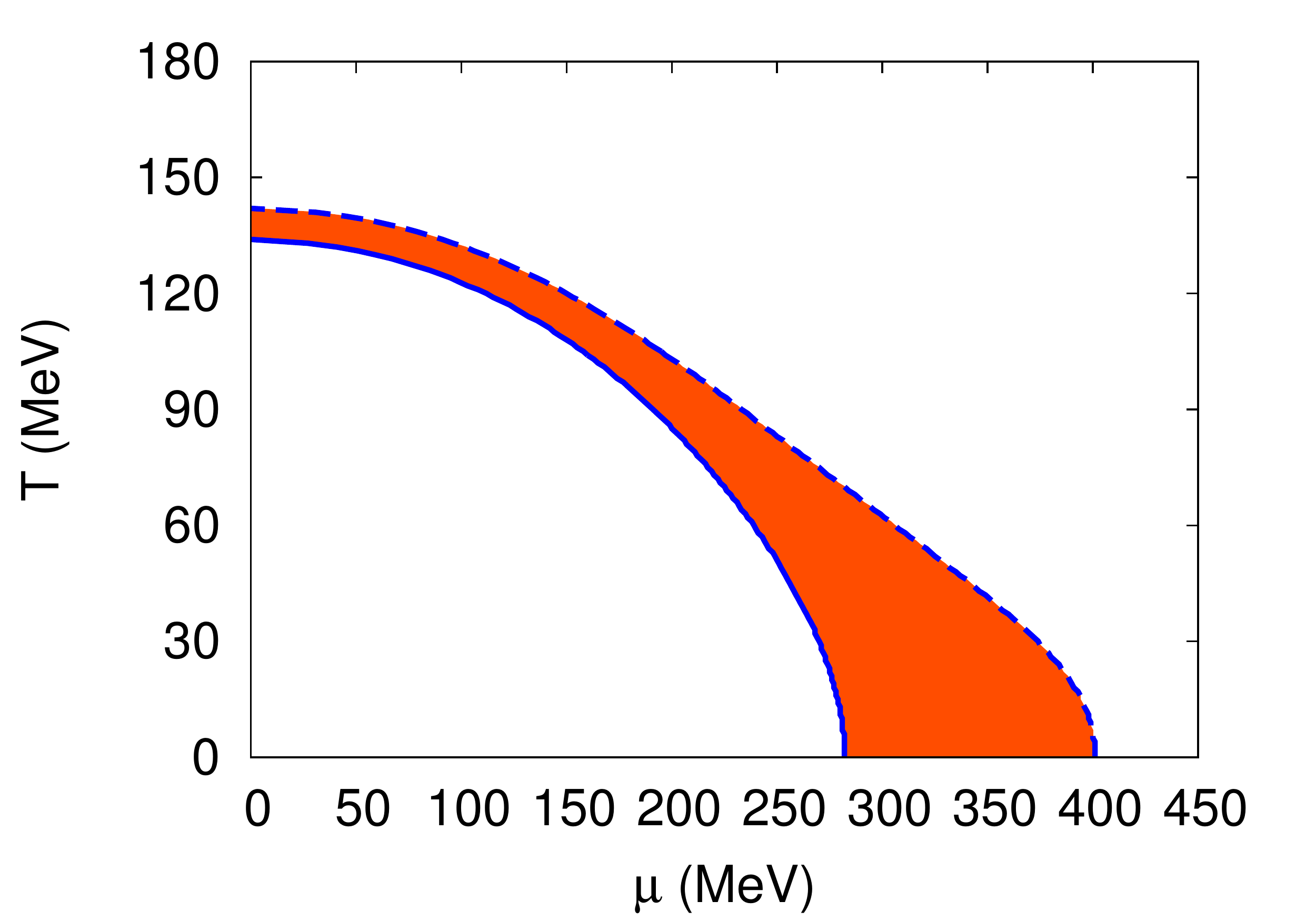}
\includegraphics[angle=270,width=.32\textwidth]{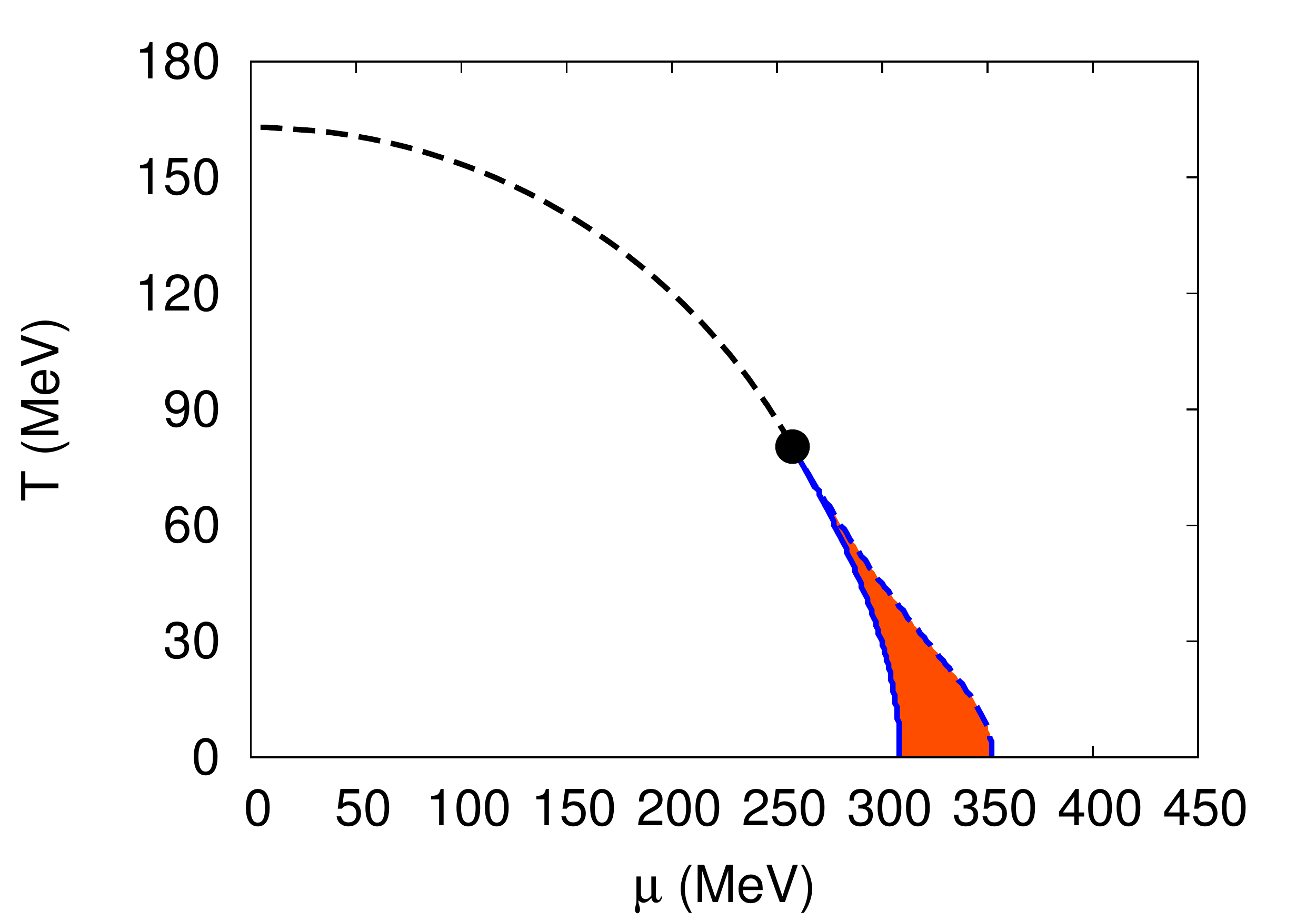}
\includegraphics[angle=270,width=.32\textwidth]{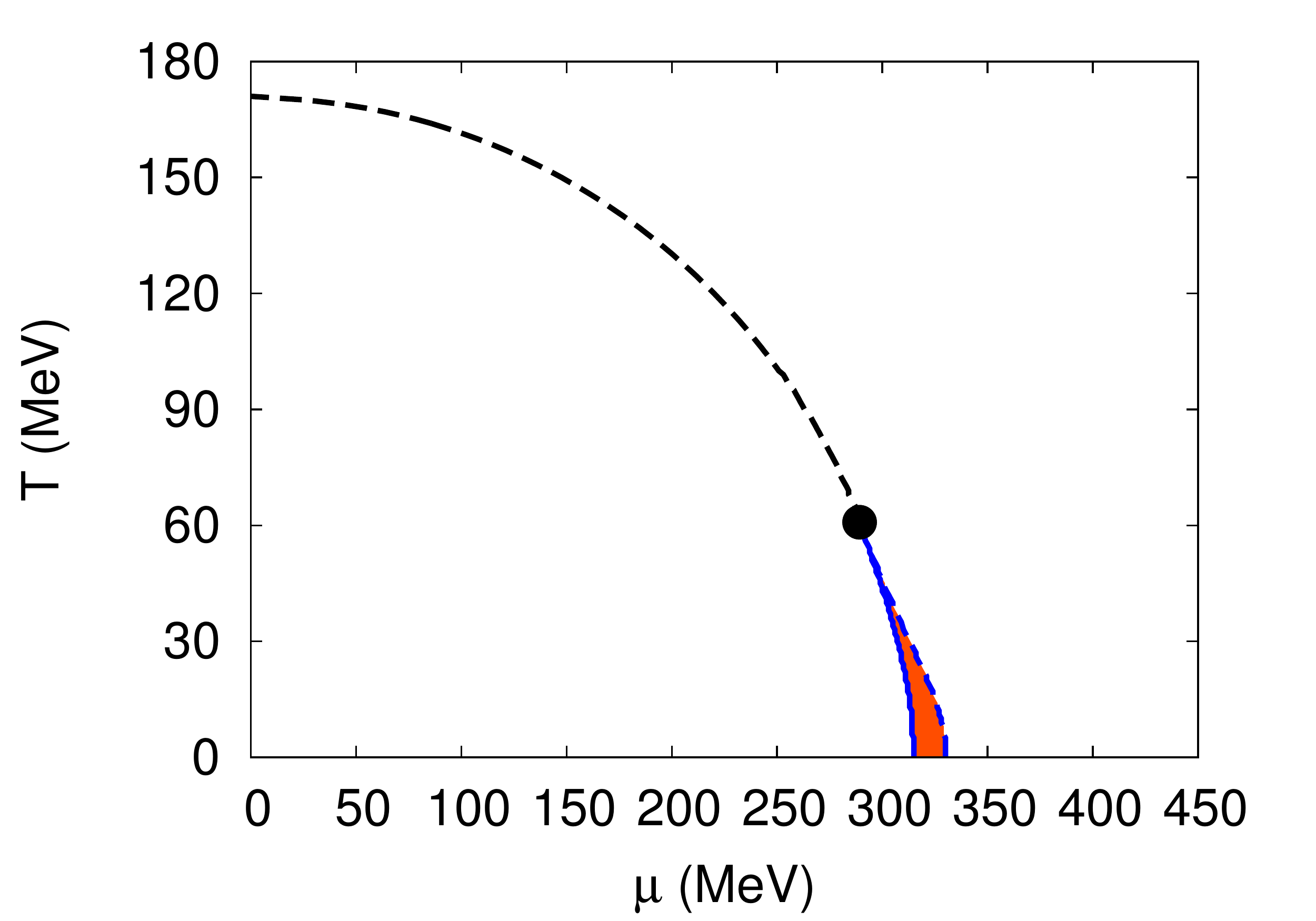}
\caption{Phase diagram, allowing for CDW modulations of the order
  parameter.  Left: results obtained in the sMFA.  Middle: including
  the Dirac sea with $\Lambda=600$~MeV.  Right: ``renormalized''
  results obtained with $\Lambda=5$~GeV.  First-order phase
  transitions are indicated by solid lines, second-order transitions
  by dashed lines.  The shaded areas indicate the inhomogeneous phase,
  the black dot denotes the Lifshitz point, coinciding with the
  location of the critical point for homogeneous phases.  }
\label{fig:pd_cdw}
\end{figure}

Turning on finite temperature and chemical potential and minimizing at
each $T$ and $\mu$ the thermodynamic potential with respect to
$\Delta$ and $q$ we obtain the phase diagrams shown in
Fig.~\ref{fig:pd_cdw}.  The three panels correspond to different
values of the PV cutoff, $\Lambda = 0$, 600~MeV and 5~GeV.  The shaded
areas indicate the CDW-type inhomogeneous regions, characterized by
nonvanishing values of $\Delta$ and $q$.  At the upper-$\mu$ side of
these regions, the amplitude $\Delta$ smoothly goes to zero,
corresponding to a second-order phase transition to the restored
phase.  At the lower-$\mu$ side, on the other hand, we find a
first-order phase transition, where an inhomogeneous solution with a
nonvanishing amplitude and $q >0$ is degenerate with a homogeneous
solution where $q =0$.

For the sigma-meson mass we have employed the NJL relation $m_\sigma =
2M$.  From the GL analysis of Sec.~\ref{sec:GL} we expect for this
case an inhomogeneous phase ending at a Lifshitz point which coincides
with the critical point of the homogeneous calculation.  Our numerical
results are in perfect agreement with this expectation.  In the sMFA,
as we have seen in Sec.~\ref{sec:pdhom}, there is no CP in the
homogeneous case.  Consequently, when we allow for CDW-type solutions,
there is no LP either, and the inhomogeneous phase extends to the
$T$-axis (left panel).  A similar result was already obtained in
Ref.~\cite{Nickel:2009wj} for a different modulation.  However, when
the Dirac sea is included (middle and right panels), the inhomogeneous
phase shrinks and a Lifshitz point appears, coinciding with the
corresponding critical point in Fig.~\ref{fig:allpdhom}.  Increasing
the cutoff reduces the size of the inhomogeneous
phase but does not destroy it, and beyond $\Lambda = 2$~ GeV its size
stays almost constant.  The renormalized QM model in extended
mean-field approximation therefore still features an inhomogeneous
phase with the same qualitative properties as the NJL model (as long
as $m_\sigma = 2 M$). In particular, the first-order phase boundary of
the homogeneous case is completely covered by the inhomogeneous phase.

In light of previous studies in the NJL model~\cite{Nickel:2009wj,
  Carignano:2010ac, Carignano:2011gr}, we expect the results on the
phase structure to be qualitatively independent of the shape of the
modulation considered.  In fact, as long as the phase transitions are
second order, the position of the LP is completely fixed by the GL
analysis of Sec.~\ref{sec:GL} and therefore independent on the
modulation.  This argument can be extended to the entire second-order
phase boundary from the inhomogeneous to the chirally restored phase.
The phase transition from the homogeneous chirally broken to the
inhomogeneous phase, on the other hand, is first order for the CDW and
therefore not universal.

For completeness, we therefore briefly consider a different kind of
modulation, namely a one-dimensional real soliton lattice of the
form~\cite{Nickel:2009wj}
\beq
M(z) = \Delta \nu \frac{sn(\Delta z, \nu) cn(\Delta z, \nu)}{dn(\Delta z, \nu)} \,,
\eeq
where $sn$, $cn$ and $dn$ are Jacobi elliptic functions with elliptic
modulus $\nu \in [0,1]$ and a dimensionful parameter $\Delta$.  This
ansatz, which has been adapted from the $1+1$ dimensional Gross-Neveu
model~\cite{Schnetz:2004vr} to the $3+1$ dimensional case, corresponds
to the most favored inhomogeneous solution known so far, both for the
NJL model and for the QM model.  It interpolates smoothly between a
single domain-wall soliton ($\nu = 1$) and a sinusoidal shape ($\nu =
0$), and allows for second-order phase transitions on both sides,
i.e., to the homogeneous broken and restored phases.  Since for this
kind of modulation the quark spectral density is also known
analytically, the formalism is identical to the one employed for the
CDW, the only difference being the function $\rho(E;\sigma,\vec\pi)$
and the spatial average of the mesonic Hamiltonian density in
\Eq{eq:mesonpotCDW}.  Details can be found in
Ref.~\cite{Nickel:2009wj}.

\begin{figure}[htp]
\centering
\includegraphics[angle=270,width=.4\textwidth]{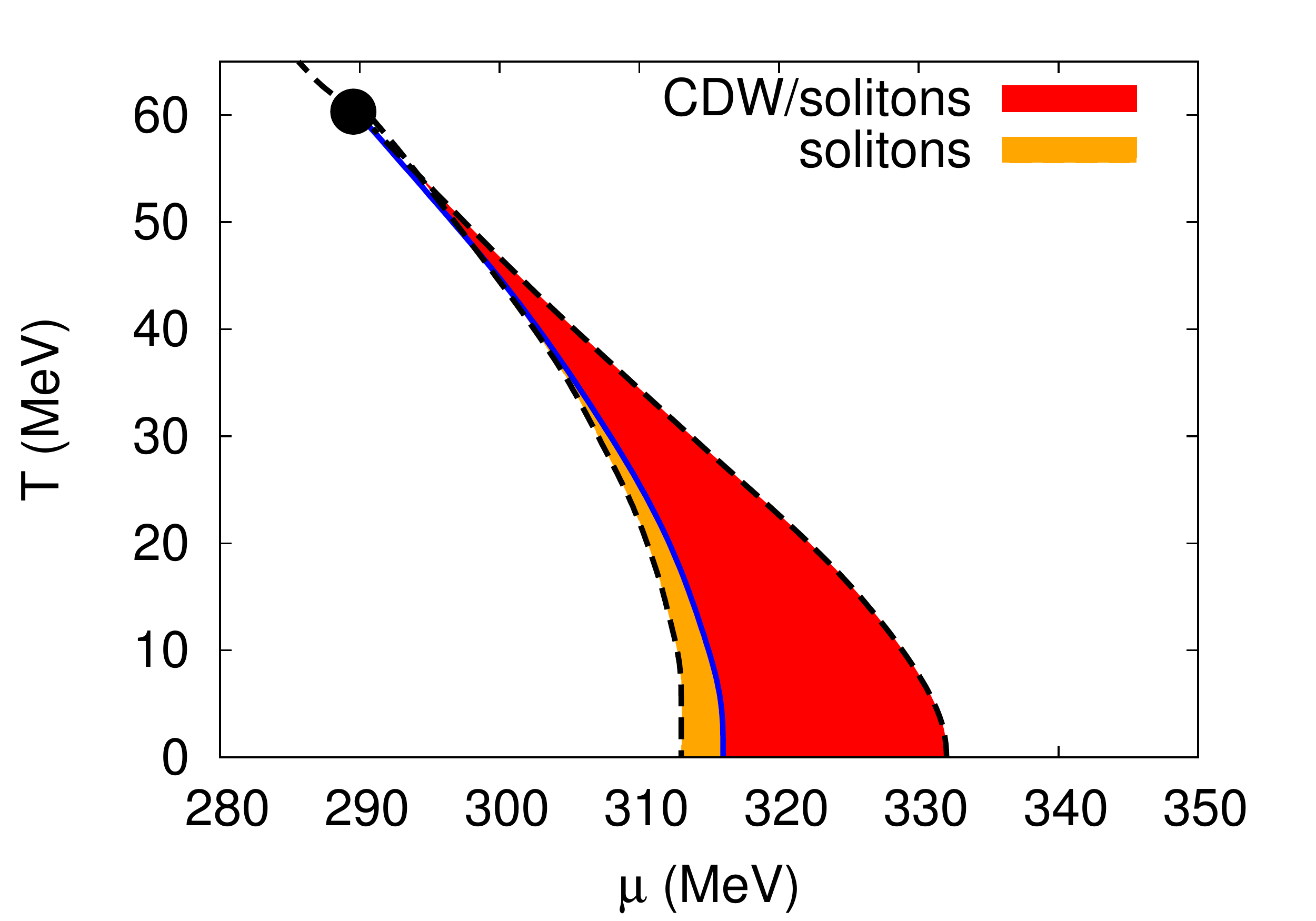}
\caption{Comparison of the renormalized ($\Lambda = 5$ GeV)
  inhomogeneous phase diagrams for the solitons and the CDW
  modulation.  The light shaded area denotes the region where only
  solitonic solutions are favored over the homogeneous
   phases. In the dark shaded area
    the CDW is favored over the homogeneous phases as well, but still
    disfavored against the solitonic modulation.   The dashed lines
  are second-order phase transitions. The solid line denotes the
  first-order phase transition at the onset of the CDW phase if
  solitonic solutions are not considered. }
\label{fig:pdsoli}
\end{figure}

In \Fig{fig:pdsoli} we compare the renormalized ($\Lambda = 5$ GeV)
phase diagram for the solitonic modulation with the one obtained for
the CDW. As expected, and just like in the NJL model, the onset of the
inhomogeneous phase slightly moves to lower chemical potential, while
the transition to the chirally restored phase and the location of the
LP\footnote{Strictly speaking, a Lifshitz point is a point where three
  second-order phase boundaries meet. This is the case for the
  solitonic modulation but not for the CDW. However, since the
  location of the point is the same, we call it LP in both cases.}
are left unchanged.  Except for the order of the phase transition
between inhomogeneous and homogeneous broken phase, no new qualitative
differences arise.

\subsection{Sensitivity of the phase structure on the sigma mass}
\label{sec:sigmamasses}

In this section we want to investigate the sensitivity of the phase
structure on the sigma mass. Empirically, the use of different input
values for $m_\sigma$ is justified since the identification of the
corresponding resonance, for example
$f_0(600)$~\cite{Beringer:1900zz}, is not yet fully established. 
Indeed, some authors favor a higher mass (above 1~GeV) for the chiral
partner of the pion and suggest that the $f_{0}$ is a tetraquark
or two-meson bound state~\cite{Pelaez:2003dy, Parganlija:2012fy}. A
strong sigma-mass sensitivity on the homogeneous phase structure,
including the existence of a CP has been observed in various two- and
three- flavor quark-meson models with and without the
Polyakov-loop~\cite{Schaefer:2008hk}.  Moreover, as we have seen in
Sec.~\ref{sec:GL}, the coincidence of the LP and the CP is special to
the NJL mass ratio, $m_\sigma /M= 2$.  Thus, choosing different ratios
offers the possibility to separate the two points.

\begin{figure}[htp]
\centering
\includegraphics[angle=270,width=.48\textwidth]{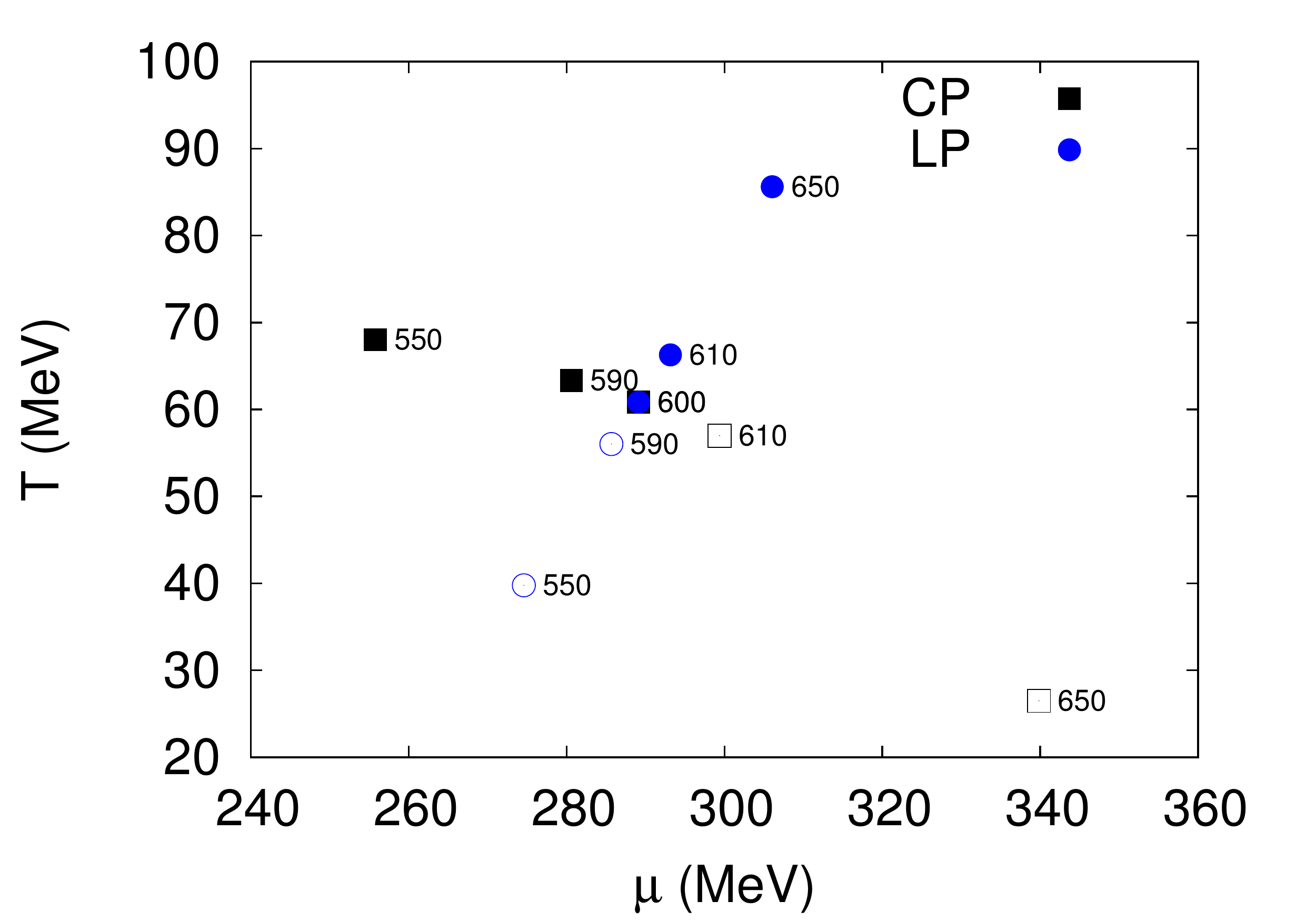}
\caption{Location of the critical (black squares) and Lifshitz (blue
  circles) points according to the Ginzburg-Landau analysis for
  $\Lambda = 5$ GeV.  The numbers indicate the corresponding values of
  the sigma mass in MeV.  The full circles are reliable, the empty
  ones are not because the analysis is invalidated by the homogeneous
  first-order phase transition.  The critical points indicated by open
  squares are only relevant if the analysis is restricted to
  homogeneous phases.}
\label{fig:CPLPmsigma}
\end{figure}

In the following, various values around $m_\sigma = 600$ MeV are
chosen while the constituent quark mass will be kept constant at
$M=300$~MeV in vacuum.  In \Fig{fig:CPLPmsigma} we show the critical
and Lifshitz points, as determined from the GL conditions
Eqs.~(\ref{eq:GLCP}) and (\ref{eq:GLLP}), for different values of the
sigma mass.  One can see that both points are rather sensitive to a
change of $m_\sigma$, and the trajectories cross each other at
$m_\sigma = 600$~MeV, as expected.  However, while both points move to
higher chemical potentials when $m_\sigma$ is increased, the CP is
shifted to lower and the LP to higher temperatures.  From this
behavior we expect that for $m_\sigma > 2M$ the CP is covered by the
inhomogeneous phase and thus disappears from the phase diagram.  For
$m_\sigma < 2M$, on the other hand, the CP is outside of the
inhomogeneous region, so that at least a part of the first-order phase
boundary between the homogeneous chirally broken and restored phases
survives.

In this context we recall that the GL analysis is only reliable for
small values of the order parameter and its derivatives.  Therefore
the prediction for the LP for $m_\sigma < 2 M$ becomes questionable,
since in this case it could lie inside the spinodal region of the
homogeneous first-order phase transition where a homogeneous solution
with large mass could be favored.

\begin{figure}[htp]
\centering
\includegraphics[angle=270,width=.48\textwidth]{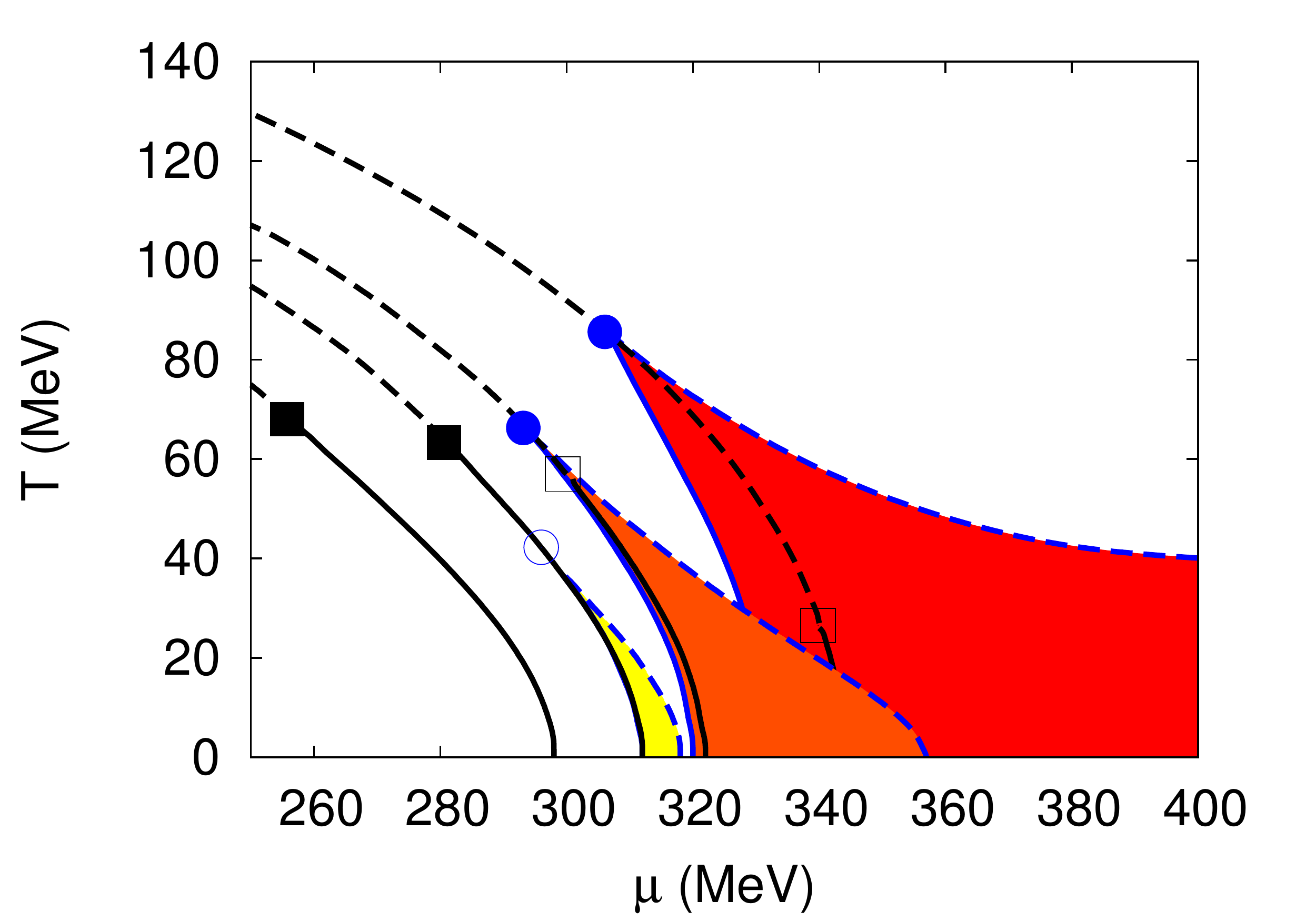}
\caption{Phase diagrams for different sigma masses, evaluated with
  $\Lambda = 5$~GeV. From left to right: $m_\sigma~=~550, 590, 610$
  and 650 MeV.  The shaded areas indicate the regions where the CDW is
  favored.  The black lines denote the phase boundaries between
  homogeneous phases (solid: first-order; dashed: second-order).
  Filled squares are the corresponding true CPs, while open squares
  indicate the locations of the CPs which would be present in a
  homogeneous analysis but are covered by the inhomogeneous phase.
  Filled circles are the LPs which agree with the predictions of the
  GL analysis, while the open circle indicates the point where the
  inhomogeneous phase ends on the first-order phase boundary between
  homogeneous phases, and which does not agree with the LP of the GL
  analysis.  }
\label{fig:pdmsigma}
\end{figure}

A full numerical study leads to the results displayed in
\Fig{fig:pdmsigma}, where we show the phase diagrams obtained for four
different values of the sigma mass.  For the inhomogeneous phase we
restrict ourselves to the CDW ansatz.  The qualitative behavior is in
good agreement with our expectations from the GL analysis, but as
explained above, there are quantitative deviations for $m_\sigma <
2M$.  In this case the inhomogeneous phase shrinks even more quickly
with decreasing $m_\sigma$ than predicted by the GL analysis, and
already at $m_\sigma = 570$~MeV it disappears completely.  Above this
value, but still below $m_\sigma = 600$~MeV, a small inhomogeneous
region exists, ending on the first-order phase boundary between the
homogeneous phases.  For $m_\sigma > 2M$, on the other hand, the size
of the inhomogeneous region is strongly enhanced, and the CP lies
inside that region.

\section{Model artifacts, islands and continents}
\label{sec:continent}

In this section we investigate in more detail the phase structure at
higher chemical potentials.  Previous studies of inhomogeneous phases
within the Nambu--Jona-Lasinio model exhibit the appearance of a
second inhomogeneous phase at high densities,
  which seems to extend up to arbitrarily high chemical
potentials~\cite{Carignano:2011gr} .  In general, such a behavior is
quite common in the literature: Inhomogeneous phases extending to
arbitrarily high chemical potentials are also found in
lower-dimensional models \cite{Schnetz:2004vr, Basar:2009fg}, in
large-$N_c$ QCD~\cite{Deryagin:1992rw, Shuster:1999tn}, in particular
in the context of quarkyonic matter~\cite{Kojo:2009ha}, as well as in
Dyson-Schwinger studies of three-color QCD~\cite{Muller:2013tya} if
color-superconductivity effects are neglected.  However, in these
systems the critical temperature of the inhomogeneous phase typically
stays constant or decreases with increasing chemical potential,
whereas in the NJL model it grows.  Hence, it is unlikely that this
inhomogeneous ``continent'' is a genuine physical feature.

Since the continent appears in a region where the chemical potential
is of the order of the NJL-model cutoff, it seems obvious that it is
caused by the regularization.  On the other hand, the formation of an
inhomogeneous phase is known to be a medium-induced
effect~\cite{Sadzikowski:2000ap, Nakano:2004cd, Kojo:2009ha}, related
to the medium term of the thermodynamic potential, which is typically
{\it not} regularized.  It has therefore been argued that the
inhomogeneous continent cannot be a trivial regularization artifact,
but rather an effect arising from the interplay between the vacuum and
the medium contributions in the thermodynamic potential
\cite{Carignano:2011gr}.  Of course, caution in the interpretation of
model results is always necessary as soon as the  chemical potential
  exceeds the model cutoff.  Hence, a similar analysis within the QM
model might help to elucidate the physical relevance of the
inhomogeneous continent.

 \begin{figure}[ht!]
\centering
\includegraphics[angle=270,width=.32\textwidth]{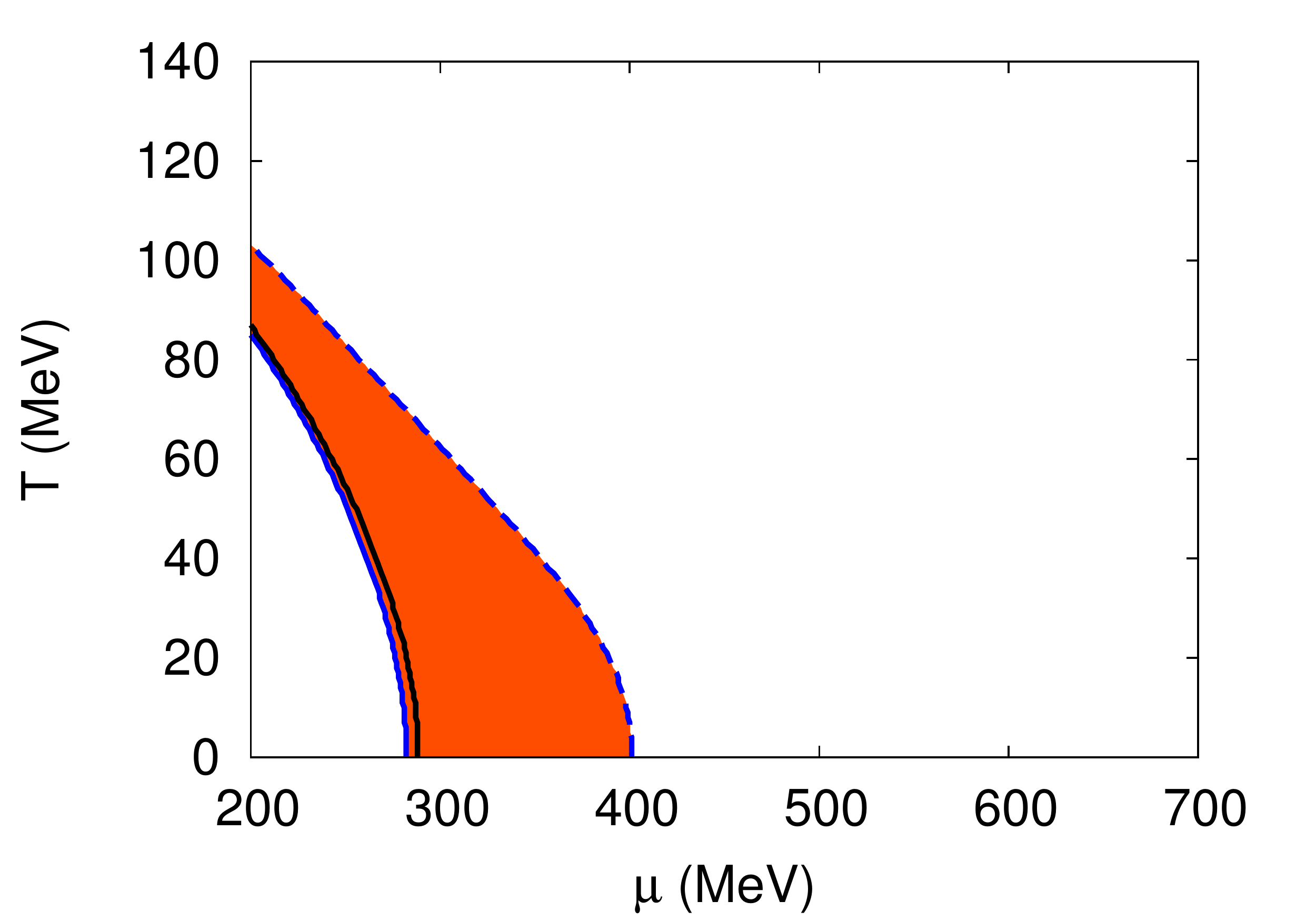}
\includegraphics[angle=270,width=.32\textwidth]{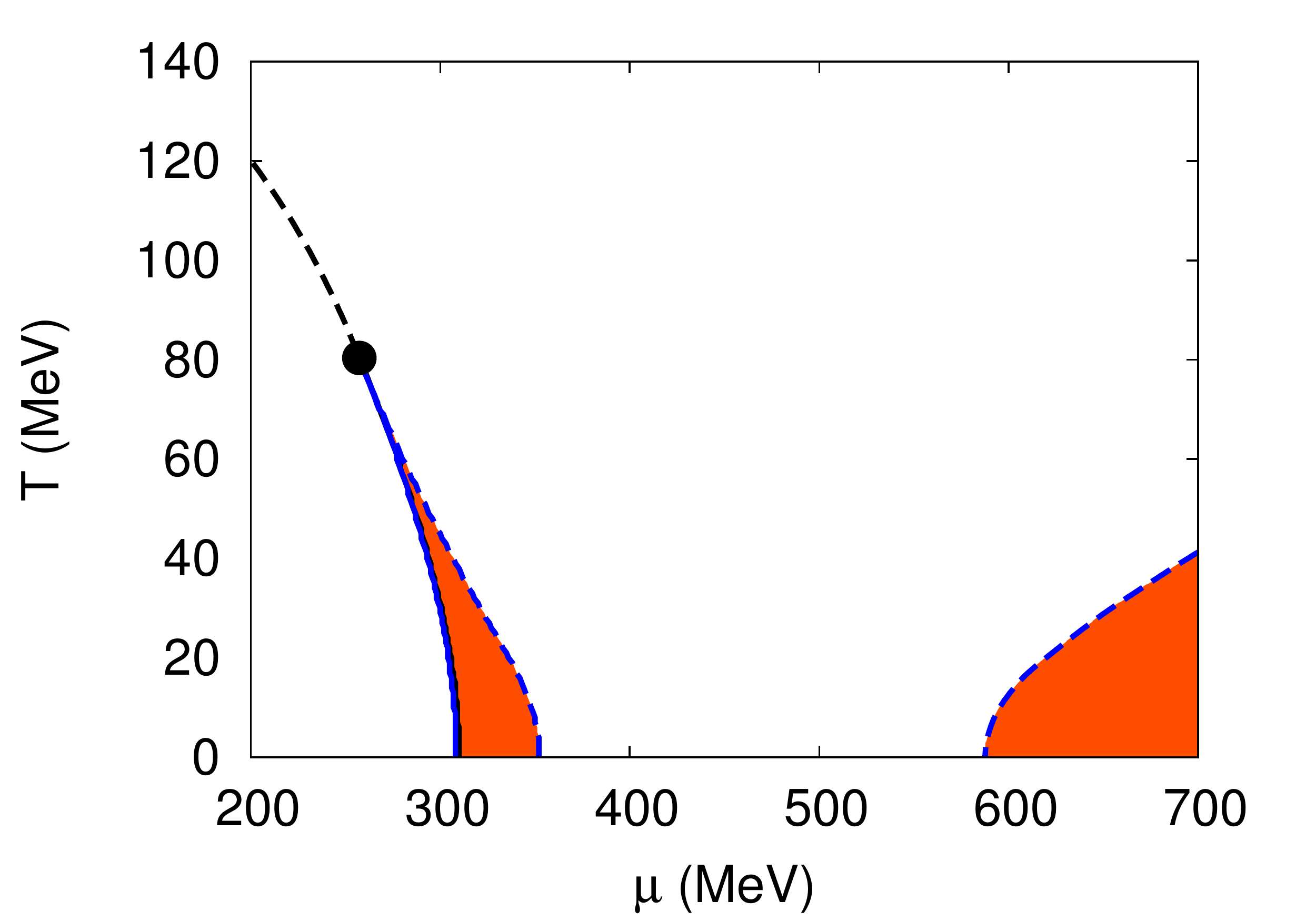}
\includegraphics[angle=270,width=.32\textwidth]{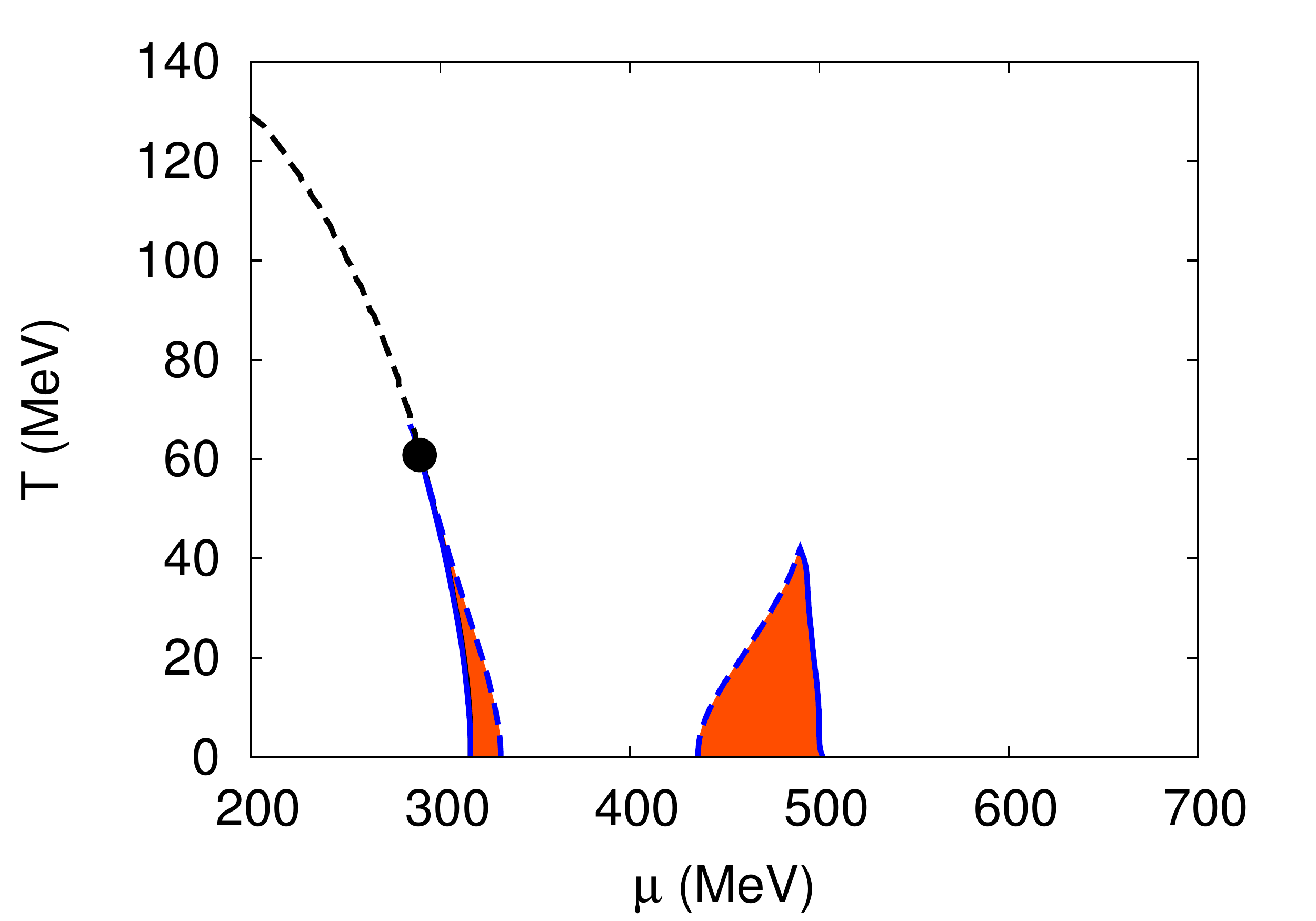}
\caption{The same phase diagrams as in Fig.~\ref{fig:pd_cdw}, extended
  to higher chemical potentials (Left: sMFA, middle: $\Lambda =
  600$~MeV, right: $\Lambda = 5$~GeV).  }
\label{fig:conti}
\end{figure}

For this purpose, we extend the phase diagrams of \Fig{fig:pd_cdw} to
higher $\mu$ and investigate the appearance of a second inhomogeneous
phase.  The results are shown in \Fig{fig:conti}, again without (left)
and with (middle and right) the Dirac-sea contribution.  As one can
see, the latter has a strong influence on the existence and the
properties of the second inhomogeneous phase.

In the left phase diagram, calculated in sMFA, no new inhomogeneous
phase appears at high densities.  The kinetic term in the meson
potential, \Eq{eq:mesonpotCDW}, disfavors solutions with high $q$,
which are characteristic of the inhomogeneous
continent~\cite{Carignano:2011gr}.  Including the vacuum contribution,
the picture changes significantly.  For an intermediate cutoff value
of, e.g., $\Lambda = 600$ MeV (middle panel of \Fig{fig:conti}), a
second inhomogeneous region appears and, just like in the NJL model,
seems to extend up to arbitrarily high values of
the chemical potential.

\begin{figure}[h]
\centering
\includegraphics[angle=270,width=.4\textwidth]{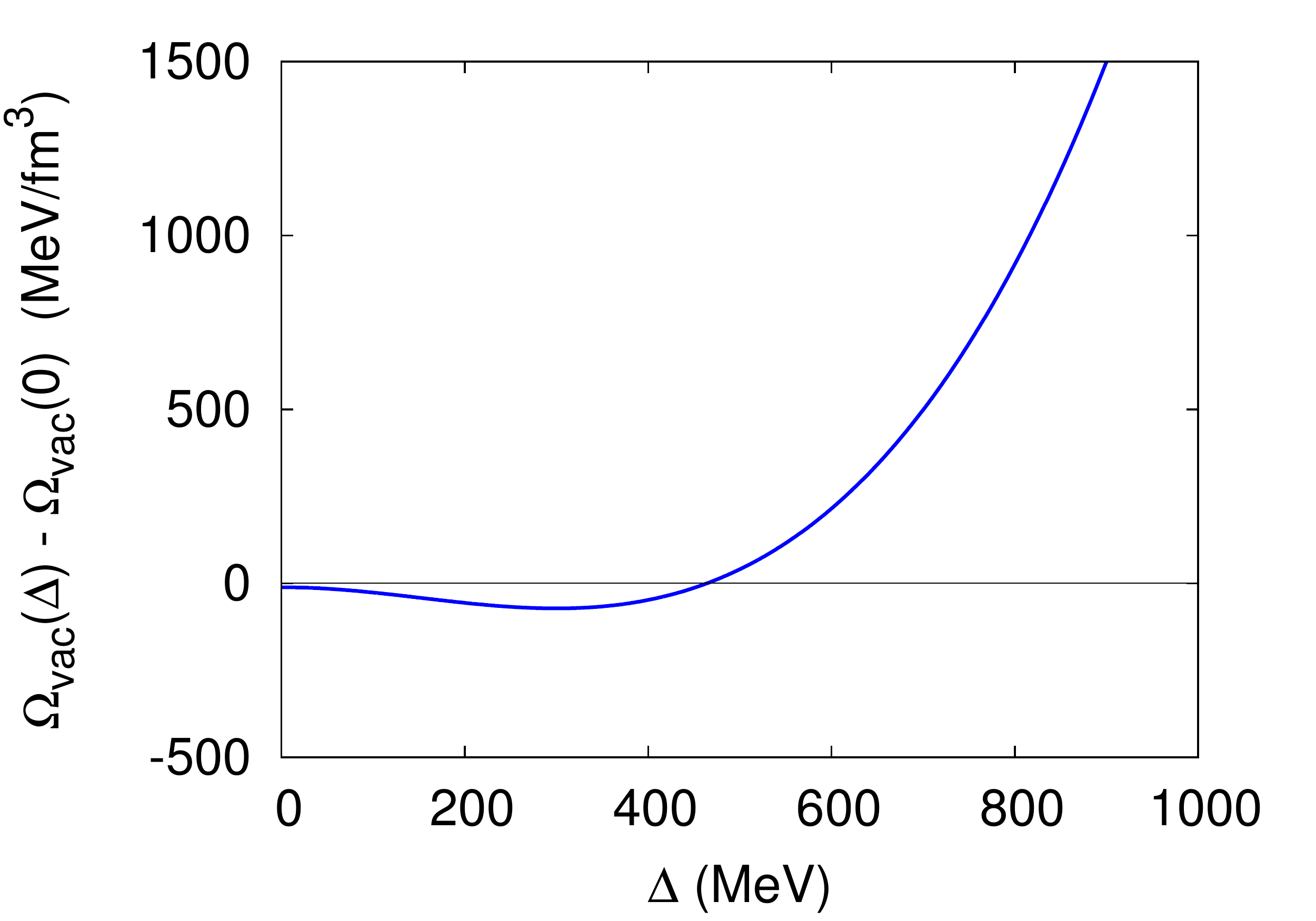}
\includegraphics[angle=270,width=.4\textwidth]{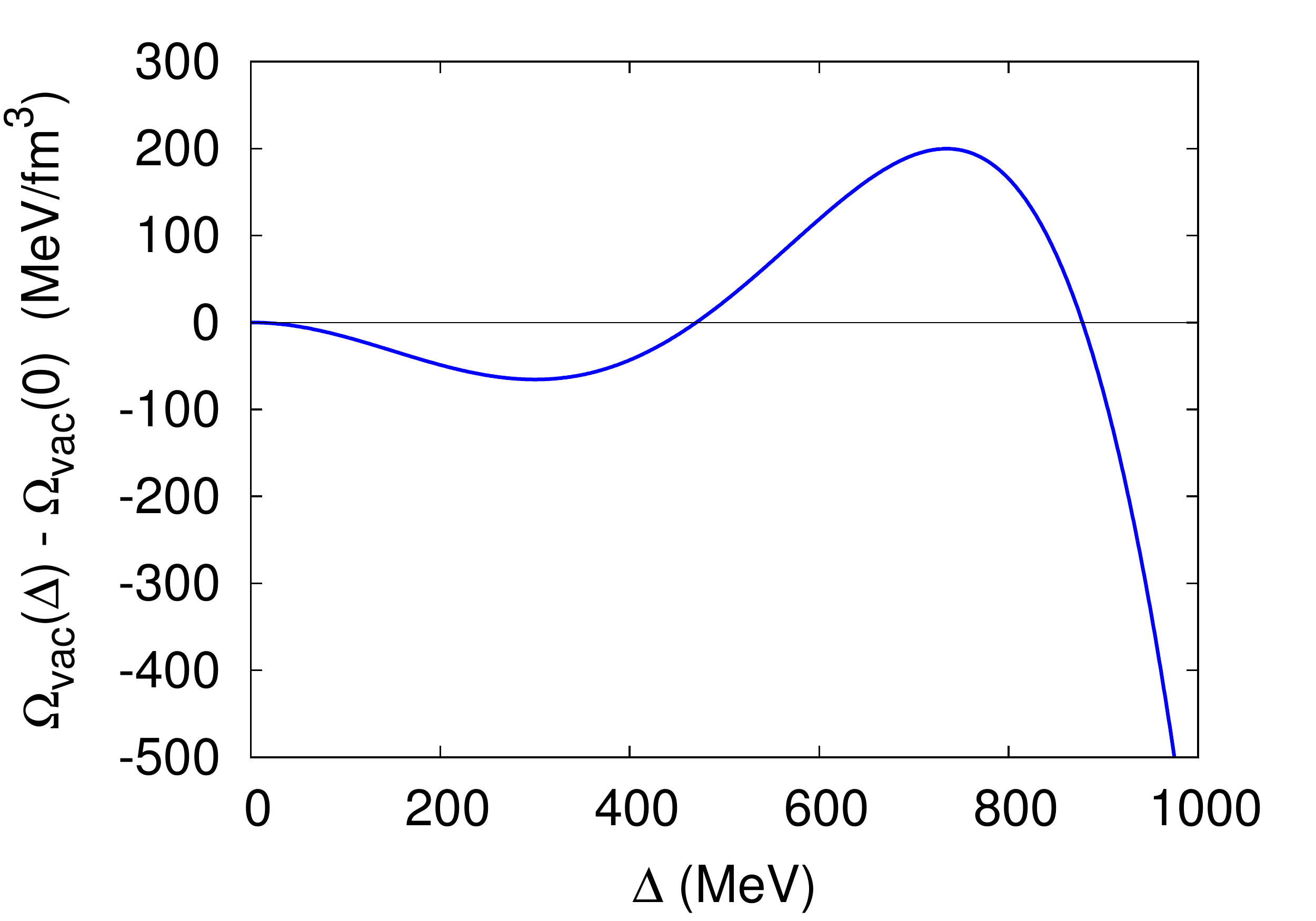}
\caption{Thermodynamic potential $\Omega_\text{vac}(\Delta) \equiv
  \Omega_\text{vac}(\sigma = \frac{\Delta}{g}, \vec\pi = 0)$ for the
  homogeneous vacuum relative to the free energy of the chirally
  restored vacuum, $\Omega_\text{vac}(0)$, as a function of the order
  parameter $\Delta$ for $\Lambda = 600$~MeV (left) and $5$~GeV
  (right).  }
\label{fig:omegadipM}
\end{figure}

Again a completely different behavior arises when the Pauli-Villars
cutoff is increased beyond the point where the model parameters
$g^2$ and $\lambda$ diverge and change their sign, i.e., in our case 
at $\Lambda \approx 750$~MeV (cf.~Fig. 3).  In this case we still find
a second inhomogeneous phase, but restricted to a finite region.
This is shown in the right panel of Fig.~10 for $\Lambda = 5$~GeV,
where the solution disappears above $\mu \gtrsim 500$~MeV.  
This disappearance however is not related to a transition to a more favored
phase, but rather to an instability.
This can be seen from the mesonic contribution to the thermodynamic potential,
\Eq{eq:Omega_M_CDW}, which becomes unbounded from below with respect to 
$\Delta$ when $\lambda$  gets negative and with respect to $q$ when $g^2$ gets negative.
Instabilities with respect to the chiral order parameter
have already been observed for the homogeneous vacuum~\cite{Skokov:2010sf}.  
This is illustrated in the right panel of \Fig{fig:omegadipM},
where one can see that the thermodynamic potential is unbounded from below
at large values of $\Delta$. 
For comparison the behavior for a smaller cutoff is shown in the left panel.
In both cases, the
vacuum constituent mass, which was fixed to 300~MeV by our
renormalization procedure (cf.~Sec.~\ref{sec:parameters}), corresponds
to a minimum.  However, for $\Lambda = 600$~MeV it is a global
minimum, whereas for $\Lambda = 5$~GeV it is only a local one,
followed by a maximum beyond which the potential decreases without
bounds.  One can show that this problem is caused by the one-loop
approximation~\cite{Coleman:1973jx}, and it can be cured by going
beyond the present approximation, e.g., with non-perturbative
functional methods such as the functional renormalization group,
cf.~\cite{Schaefer:2004en}.

\begin{figure}[h]
\centering
\includegraphics[angle=270,width=.4\textwidth]{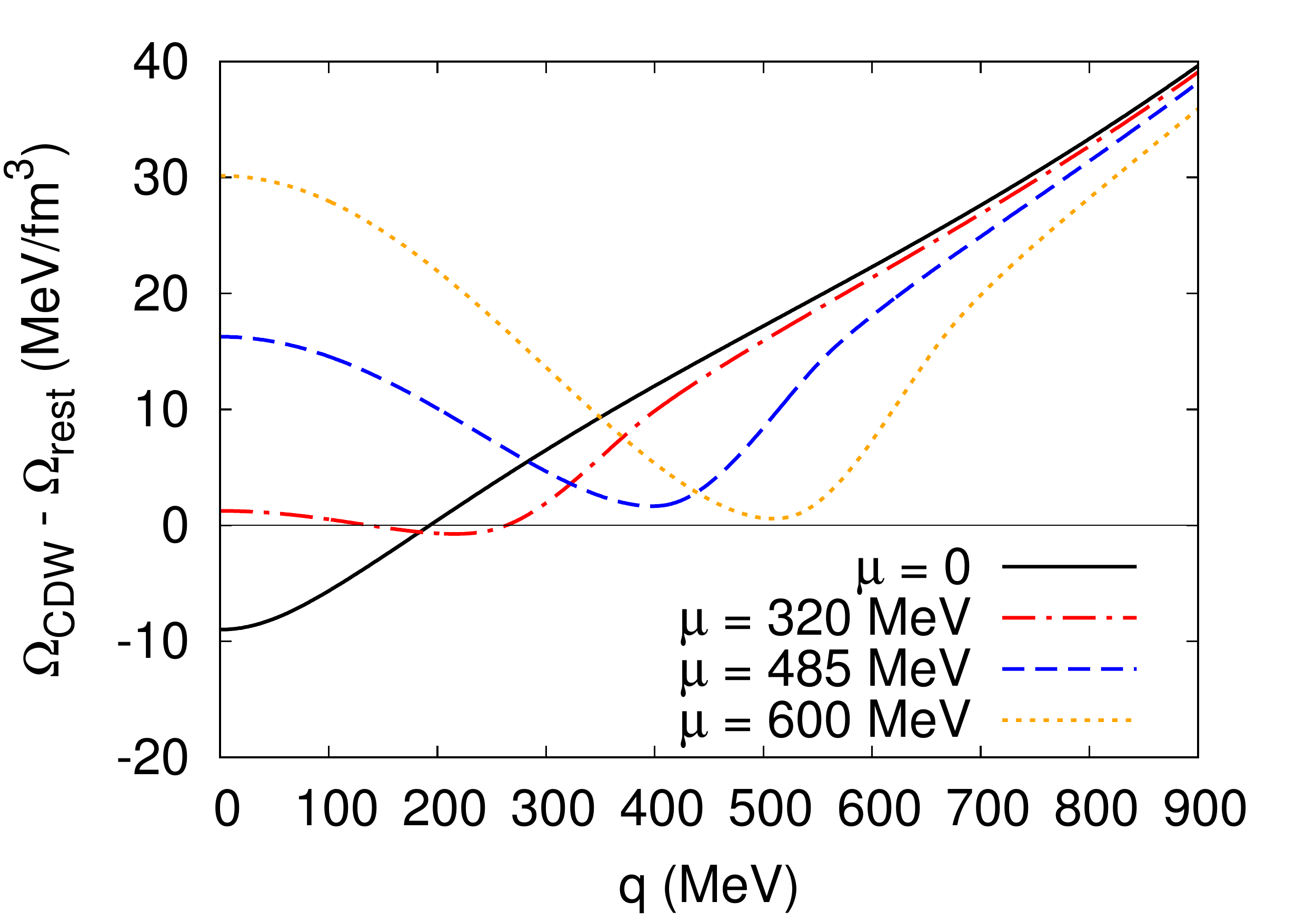}
\includegraphics[angle=270,width=.4\textwidth]{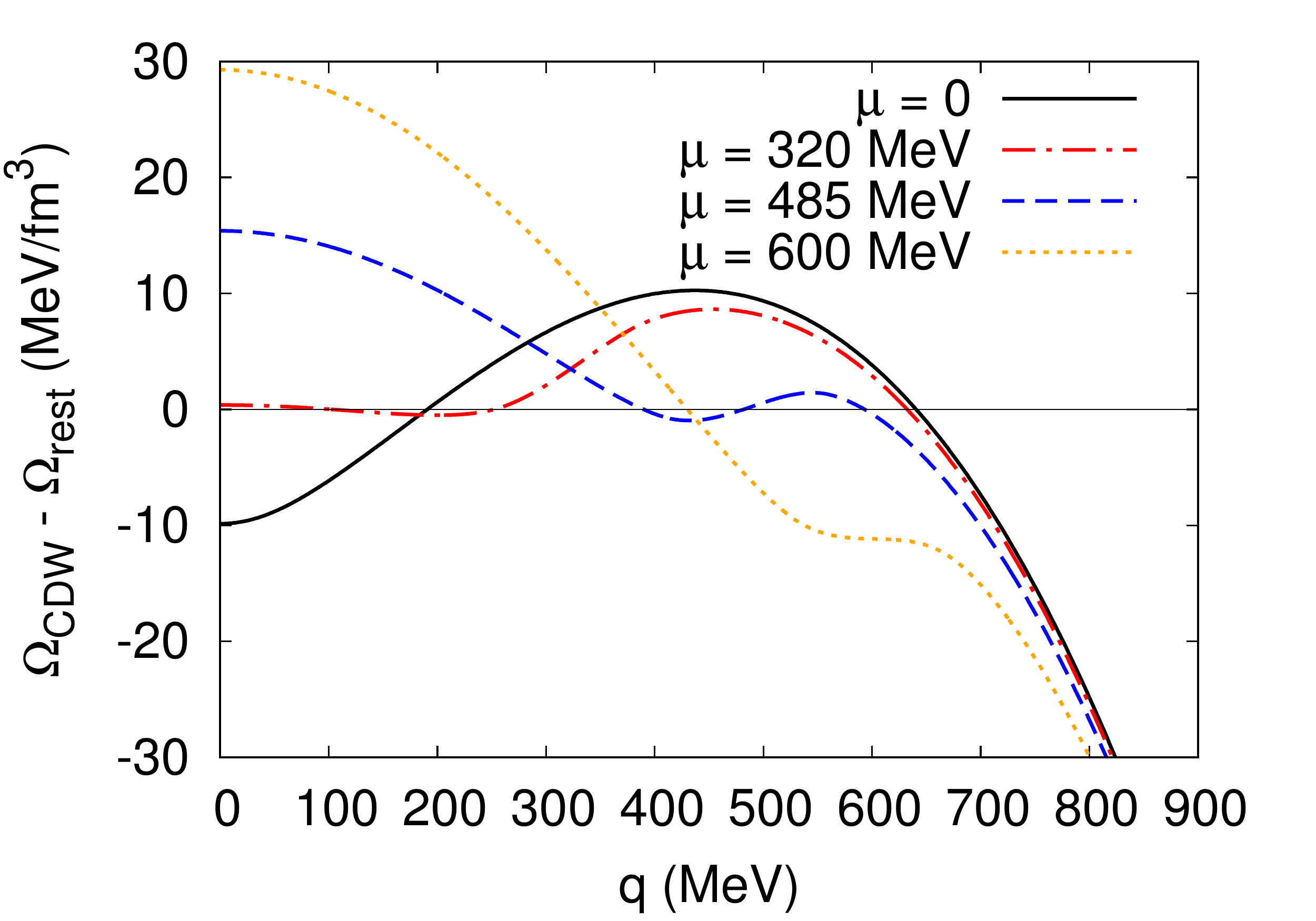}
\caption{CDW thermodynamic potential $\Omega_\text{CDW} (T=0, \mu ;
  \Delta=75 \text{ MeV}, q)$ for four different chemical potentials
  $\mu$, relative to the restored phase $\Omega_\text{rest} \equiv
  \Omega_\text{CDW} (0, \mu ; 0, 0)$ as a function of the wave number
  $q$ for $\Lambda = 600$~MeV (left) and 5 GeV (right).  }
\label{fig:omegadip}
\end{figure}

When CDW modulations are considered, a similar situation arises with
respect to the wave number.  This is demonstrated in
\Fig{fig:omegadip} where the thermodynamic potential for $T=0$ and
four different chemical potentials is plotted as a function of $q$ at
a fixed amplitude $\Delta$.  For $\Lambda = 600$~MeV (left panel) no
instability occurs, whereas for $\Lambda = 5$~GeV (right panel) the
thermodynamic potentials are again unbounded from below. In the latter
case we observe that already in vacuum (solid line) the homogeneous
chirally broken solution at $q=0$ only corresponds to a local minimum,
while arbitrarily large negative values of the thermodynamic potential
are reached at large $q$. Interestingly, a similar behavior was
already discussed more than 20 years ago in
Ref.~\cite{Broniowski:1990gb} for slightly different models.

Comparing the different curves in \Fig{fig:omegadip}, one sees that
the local minima move towards higher wave numbers with increasing
chemical potential.  The modifications of the thermodynamic potential
are essentially restricted to wave numbers of the order of $\mu$ or
less (cf. Eqs.~(\ref{eq:Omega_dos}) and (\ref{eq:rho_CDW})), i.e.,
they do not affect the behavior at high $q$.  As a consequence, the
instabilities for large cutoffs persist.  For 5~GeV (right panel),
there are still local minima for $\mu=320$ (dash-dotted line) and
$485$~MeV (dashed line), which can be related to the first and the
second inhomogeneous region in the phase diagram, respectively, (cf.\@
right panel of Fig.~\ref{fig:conti}).\footnote{As explained earlier,
  the phase diagrams are obtained by minimization of
  $\Omega_\text{CDW}$ with respect to $\Delta$ and $q$. In order to
  keep the presentation transparent, the thermodynamic potential was
  plotted for the same fixed amplitude $\Delta=75$~MeV for all curves
  shown in \Fig{fig:omegadip}. For $\mu=320$ and $485$~MeV this is
  close to the values at the local minima for $\Lambda = 5$~GeV, which
  are given by $\Delta = 82$ and $74$~MeV, respectively.}  Eventually,
however, the minimum joins with the maximum and disappears.  At
$\mu=600$~MeV (dotted line) there is therefore only an inflection
point while the local minimum is gone.

Hence, we failed to clarify the situation at large chemical
potentials: While in the NJL model (or in the QM model with a moderate
cutoff), the results are obscured by possible cutoff artifacts, in the
``renormalized'' version of the QM model they are spoiled by
unphysical instabilities.  We thus conclude that an approach beyond
the present approximation is needed in order to resolve this problem.

\section{Conclusions and outlook}
\label{sec:conclusions}

We have investigated inhomogeneous chiral-symmetry breaking phases at
non-vanishing chemical potential and temperature within a two-flavor
quark-meson model in the chiral limit.  By comparing two different
approximations of the thermodynamic potential the role of quantum
fluctuations was addressed.  We mainly focus on the effect of the
quark Dirac sea, whose importance has recently been pointed out
in~\cite{Skokov:2010sf} for homogeneous phases, but which had so far
been neglected in QM-model studies of inhomogeneous
phases~\cite{Nickel:2009wj}.  In order to include these contributions
to the thermodynamic potential, we introduced a consistent
renormalization prescription.  For the vacuum loops we employ the
Pauli-Villars regularization and fix at a given value of the
corresponding cutoff $\Lambda$ the vacuum values of the constituent
quark mass $M$, the pole mass of the sigma meson $m_\sigma$ and the
pion decay constant $f_\pi$.  By increasing the cutoff $\Lambda$ the
results converge and already for $\Lambda \gtrsim 2$~GeV no numerical
deviation is seen anymore both for homogeneous and inhomogeneous
phases.  In this way the model is effectively renormalized.  For
homogeneous phases the same asymptotic results are obtained if a sharp
$O(3)$ cutoff regularization in momentum space is applied, which is,
however, not suitable for inhomogeneous phases~\cite{Nickel:2009ke}.

Within a Ginzburg-Landau analysis we have shown that for $m_\sigma =
2M$ the Lifshitz point of the inhomogeneous phase coincides with the
tricritical point.  Remarkably, this exactly corresponds to the
NJL-model case where in the chiral limit $m_\sigma$ always equals
$2M$~\cite{Klevansky:1992qe} and the LP always coincides with the CP
\cite{Nickel:2009ke} (as long as no vector interactions are
included~\cite{Carignano:2010ac}).  In the QM model, however,
$m_\sigma$ can be varied independently from $M$, leading to scenarios
where the LP and the CP are separated.  These Ginzburg-Landau
predictions are confirmed by our numerical results where we have
considered one-dimensional modulations of the chiral order parameter.
When the Dirac sea is included the inhomogeneous phase shrinks but in
general does not disappear.  The size of the inhomogeneous region is
highly sensitive to the sigma mass and grows with increasing
$m_\sigma$.  As a consequence, for $m_\sigma > 2M$ it covers the
entire first-order phase boundary between the homogeneous phases,
including the CP, which thus disappears from the phase diagram.  For
$m_\sigma < 2M$, on the other hand, the inhomogeneous region covers at
most the low-temperature part of the homogeneous first-order line or
even vanishes completely.

Finally, we have explored the phase structure at high chemical
potentials.  For a moderate value of the cutoff, e.g., $\Lambda =
600$~MeV, an inhomogeneous ``continent'' appears, which is known from
NJL-model investigations~\cite{Carignano:2011gr} but is not present in
the sMFA.  However, for large values of the cutoff, and in particular
for the renormalized model, a general instability with respect to
both, large amplitudes and large wave numbers of the chiral
modulations are found. These numerical findings, in particular the
unphysical instabilities, might be related to the used approximation.
An improvement could be achieved, for example, via the functional
renormalization group method, which has successfully been applied to
the QM model for homogeneous phases \cite{Schaefer:2006sr} but not yet
for inhomogeneous phases.  In this way the influence of the mesonic
quantum and thermal fluctuations on the thermodynamic potential can be
analyzed in a systematic manner, in addition to the fermionic ones
which have been studied in the present work. In general, it is known
that the mesonic fluctuations are important not only in the vicinity
of a phase transition but also to stabilize the system. On the other
hand, inhomogeneous phases with one-dimensional modulations are
expected to be unstable at finite temperature~\cite{Baym:1982ca,
  Landau1969}, since the rigidity of the modulation is destroyed by
thermal fluctuations.  It would be interesting to see whether these
effects can be studied straightforwardly by including mesonic
fluctuations or whether additional degrees of freedom, in particular
phonons, must be considered.

\subsection*{Acknowledgments}

This work has been supported in part by the Helmholtz International
Center for FAIR within the LOEWE program of the State of Hesse and by
the Helmholtz Institute EMMI.  B.-J.S. acknowledges support by the FWF
Grant P24780-N27.

\appendix

\section{Dirac-sea contribution to the vacuum thermodynamic potential for small wave numbers}

Inserting the density of states, \Eq{eq:rho_CDW}, for $\Delta = M$ and $q<M$ into 
\Eq{eq:deltaomegasea}, one obtains
\bea
       \Delta\Omega_\text{vac,$q\bar q$}(q) 
       &=& 
      - \frac{N_fN_c}{2\pi^2}\int_0^\infty\!\! dE\, E\, \Big\{
       \theta(E-q-M) \sqrt{(E-q)^2 -M^2}+
\nonumber \\ 
&&
\phantom{- \frac{N_fN_c}{2\pi^2}\int_0^\infty\!\! dE\, E\, \Big\{ }
\theta(E+q-M) \sqrt{(E+q)^2 -M^2}  -  
\nonumber \\[1mm] 
&&
\phantom{- \frac{N_fN_c}{2\pi^2}\int_0^\infty\!\! dE\, E\, \Big\{ }
 2\theta(E-M) \sqrt{E^2 -M^2}  \ \Big\}  
\;\sum_j c_j \sqrt{E^2 + j\Lambda^2} 
\nonumber \\
&= &
  - \frac{N_fN_c}{2\pi^2}\int_M^\infty\!\! dE\, \sqrt{E^2-M^2}\;\sum_j c_j \, 
  \Big\{ (E+q) \sqrt{(E+q)^2 + j\Lambda^2} +
\nonumber \\
&&\phantom{- \frac{N_fN_c}{2\pi^2}\int_M^\infty\!\! dE\, \sqrt{E^2-M^2}\;\sum_j c_j \, 
  \Big\{ }
   (E-q) \sqrt{(E-q)^2 + j\Lambda^2} -
\nonumber \\[1mm]
&&\phantom{- \frac{N_fN_c}{2\pi^2}\int_M^\infty\!\! dE\, \sqrt{E^2-M^2}\;\sum_j c_j \, 
  \Big\{ }
  2E \sqrt{E^2 + j\Lambda^2} \ \Big\}  \,,
\eea
where in the second step we have shifted the integration variable by $\pm q$ in the first two lines.
Next, we Taylor-expand the integrand in $q$. This yields 
\beq
       \Delta\Omega_\text{vac,$q\bar q$}(q) 
       =
        - \frac{N_fN_c}{2\pi^2}\int_M^\infty\!\! dE\,E\, \sqrt{E^2-M^2}\;\sum_j c_j \, 
       \frac{2E^2 + 3j\Lambda^2}{(E^2+ j\Lambda^2)^{3/2}}\,q^2
       \,+\, \mathcal{O}(q^4) \,.
\eeq       
Substituting $p := \sqrt{E^2 - M^2}$ and comparing the result with \Eq{eq:L2def}, one finds
\bea
       \Delta\Omega_\text{vac,$q\bar q$}(q) 
       &=&
        - \frac{N_fN_c}{2\pi^2}\int_0^\infty\!\! dp\, p^2\;\sum_j c_j \, 
       \frac{2p^2 + 2M^2+ 3j\Lambda^2}{(p^2+M^2 + j\Lambda^2)^{3/2}}\,q^2
       \,+\, \mathcal{O}(q^4) 
\nonumber \\     
       &=&
       -M^2 L_2(0) q^2 
        - \frac{N_fN_c}{2\pi^2}\int_0^\infty\!\! dp\,  p^2\;\sum_j c_j \,
        \frac{2p^2 + 3M^2 + 3j\Lambda^2}{(p^2+M^2 + j\Lambda^2)^{3/2}} \,q^2
       \,+\, \mathcal{O}(q^4) \,.   
\eea       
The remaining integral can be written as a total derivative 
\beq
       \Delta\Omega_\text{vac,$q\bar q$}(q) 
       =
       -M^2 L_2(0) q^2 
        - \frac{N_fN_c}{2\pi^2}\int_0^\infty\!\! dp\;\sum_j c_j \,
         \frac{d}{dp}\frac{p^3}{\sqrt{p^2+M^2 + j\Lambda^2}} \,q^2
       \,+\, \mathcal{O}(q^4) \,,
\eeq
and, thus, vanishes in PV regularization.
Hence, we are left with
\beq
       \Delta\Omega_\text{vac,$q\bar q$}(q) 
       =
       -M^2 L_2(0) q^2 
       \,+\, \mathcal{O}(q^4) \,,
\eeq
which is just \Eq{eq:deltaomegaq}.



\begin{thebibliography}{60}
\expandafter\ifx\csname natexlab\endcsname\relax\def\natexlab#1{#1}\fi
\expandafter\ifx\csname bibnamefont\endcsname\relax
  \def\bibnamefont#1{#1}\fi
\expandafter\ifx\csname bibfnamefont\endcsname\relax
  \def\bibfnamefont#1{#1}\fi
\expandafter\ifx\csname citenamefont\endcsname\relax
  \def\citenamefont#1{#1}\fi
\providecommand{\bibinfo}[2]{#2}
\providecommand{\eprint}[2][]{\url{#2}}

\bibitem{BraunMunzinger:2009zz}
\bibinfo{author}{\bibfnamefont{P.}~\bibnamefont{Braun-Munzinger}}
  \bibnamefont{and} \bibinfo{author}{\bibfnamefont{J.}~\bibnamefont{Wambach}},
  \bibinfo{journal}{Rev.Mod.Phys.} \textbf{\bibinfo{volume}{81}},
  \bibinfo{pages}{1031} (\bibinfo{year}{2009}).

\bibitem{Fukushima:2010bq}
\bibinfo{author}{\bibfnamefont{K.}~\bibnamefont{Fukushima}} \bibnamefont{and}
  \bibinfo{author}{\bibfnamefont{T.}~\bibnamefont{Hatsuda}},
  \bibinfo{journal}{Rept.Prog.Phys.} \textbf{\bibinfo{volume}{74}},
  \bibinfo{pages}{014001} (\bibinfo{year}{2011}).

\bibitem{Aoki:2006we}
\bibinfo{author}{\bibfnamefont{Y.}~\bibnamefont{Aoki}},
  \bibinfo{author}{\bibfnamefont{G.}~\bibnamefont{Endrodi}},
  \bibinfo{author}{\bibfnamefont{Z.}~\bibnamefont{Fodor}},
  \bibinfo{author}{\bibfnamefont{S.}~\bibnamefont{Katz}}, \bibnamefont{and}
  \bibinfo{author}{\bibfnamefont{K.}~\bibnamefont{Szabo}},
  \bibinfo{journal}{Nature} \textbf{\bibinfo{volume}{443}},
  \bibinfo{pages}{675} (\bibinfo{year}{2006}).

\bibitem{Borsanyi:2012vn}
\bibinfo{author}{\bibfnamefont{S.}~\bibnamefont{Borsanyi}},
  \bibinfo{author}{\bibfnamefont{G.}~\bibnamefont{Endrodi}},
  \bibinfo{author}{\bibfnamefont{Z.}~\bibnamefont{Fodor}},
  \bibinfo{author}{\bibfnamefont{S.~D.} \bibnamefont{Katz}},
  \bibinfo{author}{\bibfnamefont{S.}~\bibnamefont{Krieg}},
  \bibnamefont{et~al.}, \bibinfo{journal}{PoS}
  \textbf{\bibinfo{volume}{LATTICE2011}}, \bibinfo{pages}{201}
  (\bibinfo{year}{2011}).

\bibitem{Ratti:2013uta}
\bibinfo{author}{\bibfnamefont{C.}~\bibnamefont{Ratti}},
  \bibinfo{author}{\bibfnamefont{S.}~\bibnamefont{Borsanyi}},
  \bibinfo{author}{\bibfnamefont{G.}~\bibnamefont{Endrodi}},
  \bibinfo{author}{\bibfnamefont{Z.}~\bibnamefont{Fodor}},
  \bibinfo{author}{\bibfnamefont{S.~D.} \bibnamefont{Katz}},
  \bibnamefont{et~al.}, \bibinfo{journal}{Nucl. Phys.}
  \textbf{\bibinfo{volume}{A904-905}}, \bibinfo{pages}{869c}
  (\bibinfo{year}{2013}).

\bibitem{Bazavov:2012kf}
\bibinfo{author}{\bibfnamefont{A.}~\bibnamefont{Bazavov}} \bibnamefont{et~al.}
  (\bibinfo{collaboration}{MILC Collaboration}), \bibinfo{journal}{PoS}
  \textbf{\bibinfo{volume}{LATTICE2012}}, \bibinfo{pages}{071}
  (\bibinfo{year}{2012}).

\bibitem{Friman:2011zz}
\bibinfo{author}{\bibfnamefont{B.}~\bibnamefont{Friman}},
  \bibinfo{author}{\bibfnamefont{C.}~\bibnamefont{H{\"o}hne}},
  \bibinfo{author}{\bibfnamefont{J.}~\bibnamefont{Knoll}},
  \bibinfo{author}{\bibfnamefont{S.}~\bibnamefont{Leupold}},
  \bibinfo{author}{\bibfnamefont{J.}~\bibnamefont{Randrup}},
  \bibnamefont{et~al.}, \bibinfo{journal}{Lect.Notes Phys.}
  \textbf{\bibinfo{volume}{814}}, \bibinfo{pages}{1} (\bibinfo{year}{2011}).

\bibitem{Nickel:2009ke}
\bibinfo{author}{\bibfnamefont{D.}~\bibnamefont{Nickel}},
  \bibinfo{journal}{Phys.Rev.Lett.} \textbf{\bibinfo{volume}{103}},
  \bibinfo{pages}{072301} (\bibinfo{year}{2009}).

\bibitem{Fulde:1964zz}
\bibinfo{author}{\bibfnamefont{P.}~\bibnamefont{Fulde}} \bibnamefont{and}
  \bibinfo{author}{\bibfnamefont{R.~A.} \bibnamefont{Ferrell}},
  \bibinfo{journal}{Phys.Rev.} \textbf{\bibinfo{volume}{135}},
  \bibinfo{pages}{A550} (\bibinfo{year}{1964}).

\bibitem{larkin:1964zz}
\bibinfo{author}{\bibfnamefont{A.}~\bibnamefont{Larkin}} \bibnamefont{and}
  \bibinfo{author}{\bibfnamefont{Y.}~\bibnamefont{Ovchinnikov}},
  \bibinfo{journal}{Zh.Eksp.Teor.Fiz.} \textbf{\bibinfo{volume}{47}},
  \bibinfo{pages}{1136} (\bibinfo{year}{1964}).

\bibitem{Overhauser:1962zz}
\bibinfo{author}{\bibfnamefont{A.}~\bibnamefont{Overhauser}},
  \bibinfo{journal}{Phys.Rev.} \textbf{\bibinfo{volume}{128}},
  \bibinfo{pages}{1437} (\bibinfo{year}{1962}).

\bibitem{PhysRevLett.103.207201}
\bibinfo{author}{\bibfnamefont{G.~J.} \bibnamefont{Conduit}},
  \bibinfo{author}{\bibfnamefont{A.~G.} \bibnamefont{Green}}, \bibnamefont{and}
  \bibinfo{author}{\bibfnamefont{B.~D.} \bibnamefont{Simons}},
  \bibinfo{journal}{Phys.Rev.Lett.} \textbf{\bibinfo{volume}{103}},
  \bibinfo{pages}{207201} (\bibinfo{year}{2009}).

\bibitem{Alford:2000ze}
\bibinfo{author}{\bibfnamefont{M.~G.} \bibnamefont{Alford}},
  \bibinfo{author}{\bibfnamefont{J.~A.} \bibnamefont{Bowers}},
  \bibnamefont{and}
  \bibinfo{author}{\bibfnamefont{K.}~\bibnamefont{Rajagopal}},
  \bibinfo{journal}{Phys.Rev.} \textbf{\bibinfo{volume}{D63}},
  \bibinfo{pages}{074016} (\bibinfo{year}{2001}).

\bibitem{Anglani:2013gfu}
\bibinfo{author}{\bibfnamefont{R.}~\bibnamefont{Anglani}},
  \bibinfo{author}{\bibfnamefont{R.}~\bibnamefont{Casalbuoni}},
  \bibinfo{author}{\bibfnamefont{M.}~\bibnamefont{Ciminale}},
  \bibinfo{author}{\bibfnamefont{R.}~\bibnamefont{Gatto}},
  \bibinfo{author}{\bibfnamefont{N.}~\bibnamefont{Ippolito}},
  \bibnamefont{et~al.}, \href{http://arxiv.org/abs/1302.4264}{{\tt
  arXiv:1302.4264 [hep-ph]}}.

\bibitem{Dautry:1979bk}
\bibinfo{author}{\bibfnamefont{F.}~\bibnamefont{Dautry}} \bibnamefont{and}
  \bibinfo{author}{\bibfnamefont{E.}~\bibnamefont{Nyman}},
  \bibinfo{journal}{Nucl.Phys.} \textbf{\bibinfo{volume}{A319}},
  \bibinfo{pages}{323} (\bibinfo{year}{1979}).

\bibitem{Goldhaber:1987pb}
\bibinfo{author}{\bibfnamefont{A.~S.} \bibnamefont{Goldhaber}}
  \bibnamefont{and} \bibinfo{author}{\bibfnamefont{N.}~\bibnamefont{Manton}},
  \bibinfo{journal}{Phys.Lett.} \textbf{\bibinfo{volume}{B198}},
  \bibinfo{pages}{231} (\bibinfo{year}{1987}).

\bibitem{Heinz:2013hza}
\bibinfo{author}{\bibfnamefont{A.}~\bibnamefont{Heinz}},
  \bibinfo{author}{\bibfnamefont{F.}~\bibnamefont{Giacosa}}, \bibnamefont{and}
  \bibinfo{author}{\bibfnamefont{D.~H.} \bibnamefont{Rischke}},
  \href{http://arxiv.org/abs/1312.3244}{{\tt arXiv:1312.3244 [nucl-th]}}.

\bibitem{Broniowski:1990dy}
\bibinfo{author}{\bibfnamefont{W.}~\bibnamefont{Broniowski}},
  \bibinfo{author}{\bibfnamefont{A.}~\bibnamefont{Kotlorz}}, \bibnamefont{and}
  \bibinfo{author}{\bibfnamefont{M.}~\bibnamefont{Kutschera}},
  \bibinfo{journal}{Acta Phys.Polon.} \textbf{\bibinfo{volume}{B22}},
  \bibinfo{pages}{145} (\bibinfo{year}{1991}).

\bibitem{Broniowski:1990gb}
\bibinfo{author}{\bibfnamefont{W.}~\bibnamefont{Broniowski}} \bibnamefont{and}
  \bibinfo{author}{\bibfnamefont{M.}~\bibnamefont{Kutschera}},
  \bibinfo{journal}{Phys.Lett.} \textbf{\bibinfo{volume}{B242}},
  \bibinfo{pages}{133} (\bibinfo{year}{1990}).

\bibitem{Sadzikowski:2000ap}
\bibinfo{author}{\bibfnamefont{M.}~\bibnamefont{Sadzikowski}} \bibnamefont{and}
  \bibinfo{author}{\bibfnamefont{W.}~\bibnamefont{Broniowski}},
  \bibinfo{journal}{Phys.Lett.} \textbf{\bibinfo{volume}{B488}},
  \bibinfo{pages}{63} (\bibinfo{year}{2000}).

\bibitem{Nakano:2004cd}
\bibinfo{author}{\bibfnamefont{E.}~\bibnamefont{Nakano}} \bibnamefont{and}
  \bibinfo{author}{\bibfnamefont{T.}~\bibnamefont{Tatsumi}},
  \bibinfo{journal}{Phys.Rev.} \textbf{\bibinfo{volume}{D71}},
  \bibinfo{pages}{114006} (\bibinfo{year}{2005}).

\bibitem{Nickel:2009wj}
\bibinfo{author}{\bibfnamefont{D.}~\bibnamefont{Nickel}},
  \bibinfo{journal}{Phys.Rev.} \textbf{\bibinfo{volume}{D80}},
  \bibinfo{pages}{074025} (\bibinfo{year}{2009}).

\bibitem{Carignano:2010ac}
\bibinfo{author}{\bibfnamefont{S.}~\bibnamefont{Carignano}},
  \bibinfo{author}{\bibfnamefont{D.}~\bibnamefont{Nickel}}, \bibnamefont{and}
  \bibinfo{author}{\bibfnamefont{M.}~\bibnamefont{Buballa}},
  \bibinfo{journal}{Phys.Rev.} \textbf{\bibinfo{volume}{D82}},
  \bibinfo{pages}{054009} (\bibinfo{year}{2010}).

\bibitem{Carignano:2011gr}
\bibinfo{author}{\bibfnamefont{S.}~\bibnamefont{Carignano}} \bibnamefont{and}
  \bibinfo{author}{\bibfnamefont{M.}~\bibnamefont{Buballa}},
  \bibinfo{journal}{Acta Phys.Polon.Supp.} \textbf{\bibinfo{volume}{5}},
  \bibinfo{pages}{641} (\bibinfo{year}{2012}).

\bibitem{Fukushima:2012mz}
\bibinfo{author}{\bibfnamefont{K.}~\bibnamefont{Fukushima}},
  \bibinfo{journal}{Phys.Rev.} \textbf{\bibinfo{volume}{D86}},
  \bibinfo{pages}{054002} (\bibinfo{year}{2012}).

\bibitem{Carignano:2012sx}
\bibinfo{author}{\bibfnamefont{S.}~\bibnamefont{Carignano}} \bibnamefont{and}
  \bibinfo{author}{\bibfnamefont{M.}~\bibnamefont{Buballa}},
  \bibinfo{journal}{Phys.Rev.} \textbf{\bibinfo{volume}{D86}},
  \bibinfo{pages}{074018} (\bibinfo{year}{2012}).

\bibitem{Muller:2013tya}
\bibinfo{author}{\bibfnamefont{D.}~\bibnamefont{M{\"u}ller}},
  \bibinfo{author}{\bibfnamefont{M.}~\bibnamefont{Buballa}}, \bibnamefont{and}
  \bibinfo{author}{\bibfnamefont{J.}~\bibnamefont{Wambach}},
  \bibinfo{journal}{Phys.Lett.} \textbf{\bibinfo{volume}{B727}},
  \bibinfo{pages}{240} (\bibinfo{year}{2013}).

\bibitem{Tatsumi:2013nga}
\bibinfo{author}{\bibfnamefont{T.}~\bibnamefont{Tatsumi}},
  \bibinfo{author}{\bibfnamefont{K.}~\bibnamefont{Nishiyama}},
  \bibnamefont{and} \bibinfo{author}{\bibfnamefont{S.}~\bibnamefont{Karasawa}},
  \href{http://arxiv.org/abs/1312.0307}{{\tt arXiv:1312.0307 [hep-ph]}}.

\bibitem{Vogl:1991qt}
\bibinfo{author}{\bibfnamefont{U.}~\bibnamefont{Vogl}} \bibnamefont{and}
  \bibinfo{author}{\bibfnamefont{W.}~\bibnamefont{Weise}},
  \bibinfo{journal}{Prog.Part.Nucl.Phys.} \textbf{\bibinfo{volume}{27}},
  \bibinfo{pages}{195} (\bibinfo{year}{1991}).

\bibitem{Klevansky:1992qe}
\bibinfo{author}{\bibfnamefont{S.}~\bibnamefont{Klevansky}},
  \bibinfo{journal}{Rev.Mod.Phys.} \textbf{\bibinfo{volume}{64}},
  \bibinfo{pages}{649} (\bibinfo{year}{1992}).

\bibitem{Hatsuda:1994pi}
\bibinfo{author}{\bibfnamefont{T.}~\bibnamefont{Hatsuda}} \bibnamefont{and}
  \bibinfo{author}{\bibfnamefont{T.}~\bibnamefont{Kunihiro}},
  \bibinfo{journal}{Phys.Rept.} \textbf{\bibinfo{volume}{247}},
  \bibinfo{pages}{221} (\bibinfo{year}{1994}).

\bibitem{Buballa:2003qv}
\bibinfo{author}{\bibfnamefont{M.}~\bibnamefont{Buballa}},
  \bibinfo{journal}{Phys.Rept.} \textbf{\bibinfo{volume}{407}},
  \bibinfo{pages}{205} (\bibinfo{year}{2005}).

\bibitem{Abuki:2011pf}
\bibinfo{author}{\bibfnamefont{H.}~\bibnamefont{Abuki}},
  \bibinfo{author}{\bibfnamefont{D.}~\bibnamefont{Ishibashi}},
  \bibnamefont{and} \bibinfo{author}{\bibfnamefont{K.}~\bibnamefont{Suzuki}},
  \bibinfo{journal}{Phys.Rev.} \textbf{\bibinfo{volume}{D85}},
  \bibinfo{pages}{074002} (\bibinfo{year}{2012}).

\bibitem{Deryagin:1992rw}
\bibinfo{author}{\bibfnamefont{D.}~\bibnamefont{Deryagin}},
  \bibinfo{author}{\bibfnamefont{D.~Y.} \bibnamefont{Grigoriev}},
  \bibnamefont{and} \bibinfo{author}{\bibfnamefont{V.}~\bibnamefont{Rubakov}},
  \bibinfo{journal}{Int.J.Mod.Phys.} \textbf{\bibinfo{volume}{A7}},
  \bibinfo{pages}{659} (\bibinfo{year}{1992}).

\bibitem{Shuster:1999tn}
\bibinfo{author}{\bibfnamefont{E.}~\bibnamefont{Shuster}} \bibnamefont{and}
  \bibinfo{author}{\bibfnamefont{D.}~\bibnamefont{Son}},
  \bibinfo{journal}{Nucl.Phys.} \textbf{\bibinfo{volume}{B573}},
  \bibinfo{pages}{434} (\bibinfo{year}{2000}).

\bibitem{Kojo:2009ha}
\bibinfo{author}{\bibfnamefont{T.}~\bibnamefont{Kojo}},
  \bibinfo{author}{\bibfnamefont{Y.}~\bibnamefont{Hidaka}},
  \bibinfo{author}{\bibfnamefont{L.}~\bibnamefont{McLerran}}, \bibnamefont{and}
  \bibinfo{author}{\bibfnamefont{R.~D.} \bibnamefont{Pisarski}},
  \bibinfo{journal}{Nucl.Phys.} \textbf{\bibinfo{volume}{A843}},
  \bibinfo{pages}{37} (\bibinfo{year}{2010}).

\bibitem{Scavenius:2000qd}
\bibinfo{author}{\bibfnamefont{O.}~\bibnamefont{Scavenius}},
  \bibinfo{author}{\bibfnamefont{A.}~\bibnamefont{Mocsy}},
  \bibinfo{author}{\bibfnamefont{I.}~\bibnamefont{Mishustin}},
  \bibnamefont{and} \bibinfo{author}{\bibfnamefont{D.}~\bibnamefont{Rischke}},
  \bibinfo{journal}{Phys.Rev.} \textbf{\bibinfo{volume}{C64}},
  \bibinfo{pages}{045202} (\bibinfo{year}{2001}).

\bibitem{Schaefer:2006ds}
\bibinfo{author}{\bibfnamefont{B.-J.} \bibnamefont{Schaefer}} \bibnamefont{and}
  \bibinfo{author}{\bibfnamefont{J.}~\bibnamefont{Wambach}},
  \bibinfo{journal}{Phys.Rev.} \textbf{\bibinfo{volume}{D75}},
  \bibinfo{pages}{085015} (\bibinfo{year}{2007}).

\bibitem{Skokov:2010sf}
\bibinfo{author}{\bibfnamefont{V.}~\bibnamefont{Skokov}},
  \bibinfo{author}{\bibfnamefont{B.}~\bibnamefont{Friman}},
  \bibinfo{author}{\bibfnamefont{E.}~\bibnamefont{Nakano}},
  \bibinfo{author}{\bibfnamefont{K.}~\bibnamefont{Redlich}}, \bibnamefont{and}
  \bibinfo{author}{\bibfnamefont{B.-J.} \bibnamefont{Schaefer}},
  \bibinfo{journal}{Phys.Rev.} \textbf{\bibinfo{volume}{D82}},
  \bibinfo{pages}{034029} (\bibinfo{year}{2010}).

\bibitem{Schaefer:2011ex}
\bibinfo{author}{\bibfnamefont{B.-J.} \bibnamefont{Schaefer}} \bibnamefont{and}
  \bibinfo{author}{\bibfnamefont{M.}~\bibnamefont{Wagner}},
  \bibinfo{journal}{Phys.Rev.} \textbf{\bibinfo{volume}{D85}},
  \bibinfo{pages}{034027} (\bibinfo{year}{2012}).

\bibitem{Andersen:2011pr}
\bibinfo{author}{\bibfnamefont{J.~O.} \bibnamefont{Andersen}},
  \bibinfo{author}{\bibfnamefont{R.}~\bibnamefont{Khan}}, \bibnamefont{and}
  \bibinfo{author}{\bibfnamefont{L.~T.} \bibnamefont{Kyllingstad}},
  \href{http://arxiv.org/abs/1102.2779}{{\tt arXiv:1102.2779 [hep-ph]}}.

\bibitem{Gupta:2011ez}
\bibinfo{author}{\bibfnamefont{U.~S.} \bibnamefont{Gupta}} \bibnamefont{and}
  \bibinfo{author}{\bibfnamefont{V.~K.} \bibnamefont{Tiwari}},
  \bibinfo{journal}{Phys.Rev.} \textbf{\bibinfo{volume}{D85}},
  \bibinfo{pages}{014010} (\bibinfo{year}{2012}).

\bibitem{Mao:2009aq}
\bibinfo{author}{\bibfnamefont{H.}~\bibnamefont{Mao}},
  \bibinfo{author}{\bibfnamefont{J.}~\bibnamefont{Jin}}, \bibnamefont{and}
  \bibinfo{author}{\bibfnamefont{M.}~\bibnamefont{Huang}},
  \bibinfo{journal}{J.Phys.} \textbf{\bibinfo{volume}{G37}},
  \bibinfo{pages}{035001} (\bibinfo{year}{2010}).

\bibitem{Gupta:2009fg}
\bibinfo{author}{\bibfnamefont{U.~S.} \bibnamefont{Gupta}} \bibnamefont{and}
  \bibinfo{author}{\bibfnamefont{V.~K.} \bibnamefont{Tiwari}},
  \bibinfo{journal}{Phys.Rev.} \textbf{\bibinfo{volume}{D81}},
  \bibinfo{pages}{054019} (\bibinfo{year}{2010}).

\bibitem{Schaefer:2009ui}
\bibinfo{author}{\bibfnamefont{B.-J.} \bibnamefont{Schaefer}},
  \bibinfo{author}{\bibfnamefont{M.}~\bibnamefont{Wagner}}, \bibnamefont{and}
  \bibinfo{author}{\bibfnamefont{J.}~\bibnamefont{Wambach}},
  \bibinfo{journal}{Phys.Rev.} \textbf{\bibinfo{volume}{D81}},
  \bibinfo{pages}{074013} (\bibinfo{year}{2010}).

\bibitem{Chatterjee:2011jd}
\bibinfo{author}{\bibfnamefont{S.}~\bibnamefont{Chatterjee}} \bibnamefont{and}
  \bibinfo{author}{\bibfnamefont{K.~A.} \bibnamefont{Mohan}},
  \bibinfo{journal}{Phys.Rev.} \textbf{\bibinfo{volume}{D85}},
  \bibinfo{pages}{074018} (\bibinfo{year}{2012}).

\bibitem{Scadron:2013vba}
\bibinfo{author}{\bibfnamefont{M.~D.} \bibnamefont{Scadron}},
  \bibinfo{author}{\bibfnamefont{G.}~\bibnamefont{Rupp}}, \bibnamefont{and}
  \bibinfo{author}{\bibfnamefont{R.}~\bibnamefont{Delbourgo}},
  \bibinfo{journal}{Fortsch.Phys.} \textbf{\bibinfo{volume}{61}},
  \bibinfo{pages}{994} (\bibinfo{year}{2013}).

\bibitem{Jungnickel1996b}
\bibinfo{author}{\bibfnamefont{D.~U.} \bibnamefont{Jungnickel}}
  \bibnamefont{and}
  \bibinfo{author}{\bibfnamefont{C.}~\bibnamefont{Wetterich}},
  \bibinfo{journal}{Phys. Rev.} \textbf{\bibinfo{volume}{D53}},
  \bibinfo{pages}{5142} (\bibinfo{year}{1996}).

\bibitem{Strodthoff:2011tz}
\bibinfo{author}{\bibfnamefont{N.}~\bibnamefont{Strodthoff}},
  \bibinfo{author}{\bibfnamefont{B.-J.} \bibnamefont{Schaefer}},
  \bibnamefont{and} \bibinfo{author}{\bibfnamefont{L.}~\bibnamefont{von
  Smekal}}, \bibinfo{journal}{Phys.Rev.} \textbf{\bibinfo{volume}{D85}},
  \bibinfo{pages}{074007} (\bibinfo{year}{2012}).

\bibitem{Schaefer:2004en}
\bibinfo{author}{\bibfnamefont{B.-J.} \bibnamefont{Schaefer}} \bibnamefont{and}
  \bibinfo{author}{\bibfnamefont{J.}~\bibnamefont{Wambach}},
  \bibinfo{journal}{Nucl.Phys.} \textbf{\bibinfo{volume}{A757}},
  \bibinfo{pages}{479} (\bibinfo{year}{2005}).

\bibitem{Basar:2009fg}
\bibinfo{author}{\bibfnamefont{G.}~\bibnamefont{Basar}},
  \bibinfo{author}{\bibfnamefont{G.~V.} \bibnamefont{Dunne}}, \bibnamefont{and}
  \bibinfo{author}{\bibfnamefont{M.}~\bibnamefont{Thies}},
  \bibinfo{journal}{Phys.Rev.} \textbf{\bibinfo{volume}{D79}},
  \bibinfo{pages}{105012} (\bibinfo{year}{2009}).

\bibitem{Schnetz:2004vr}
\bibinfo{author}{\bibfnamefont{O.}~\bibnamefont{Schnetz}},
  \bibinfo{author}{\bibfnamefont{M.}~\bibnamefont{Thies}}, \bibnamefont{and}
  \bibinfo{author}{\bibfnamefont{K.}~\bibnamefont{Urlichs}},
  \bibinfo{journal}{Annals Phys.} \textbf{\bibinfo{volume}{314}},
  \bibinfo{pages}{425} (\bibinfo{year}{2004}).

\bibitem{Beringer:1900zz}
\bibinfo{author}{\bibfnamefont{J.}~\bibnamefont{Beringer}} \bibnamefont{et~al.}
  (\bibinfo{collaboration}{Particle Data Group}), \bibinfo{journal}{Phys.Rev.}
  \textbf{\bibinfo{volume}{D86}}, \bibinfo{pages}{010001}
  (\bibinfo{year}{2012}).

\bibitem{Pelaez:2003dy}
\bibinfo{author}{\bibfnamefont{J.}~\bibnamefont{Pelaez}},
  \bibinfo{journal}{Phys.Rev.Lett.} \textbf{\bibinfo{volume}{92}},
  \bibinfo{pages}{102001} (\bibinfo{year}{2004}).

\bibitem{Parganlija:2012fy}
\bibinfo{author}{\bibfnamefont{D.}~\bibnamefont{Parganlija}},
  \bibinfo{author}{\bibfnamefont{P.}~\bibnamefont{Kovacs}},
  \bibinfo{author}{\bibfnamefont{G.}~\bibnamefont{Wolf}},
  \bibinfo{author}{\bibfnamefont{F.}~\bibnamefont{Giacosa}}, \bibnamefont{and}
  \bibinfo{author}{\bibfnamefont{D.~H.} \bibnamefont{Rischke}},
  \bibinfo{journal}{Phys.Rev.} \textbf{\bibinfo{volume}{D87}},
  \bibinfo{pages}{014011} (\bibinfo{year}{2013}).

\bibitem{Schaefer:2008hk}
\bibinfo{author}{\bibfnamefont{B.-J.} \bibnamefont{Schaefer}} \bibnamefont{and}
  \bibinfo{author}{\bibfnamefont{M.}~\bibnamefont{Wagner}},
  \bibinfo{journal}{Phys.Rev.} \textbf{\bibinfo{volume}{D79}},
  \bibinfo{pages}{014018} (\bibinfo{year}{2009}).

\bibitem{Coleman:1973jx}
\bibinfo{author}{\bibfnamefont{S.~R.} \bibnamefont{Coleman}} \bibnamefont{and}
  \bibinfo{author}{\bibfnamefont{E.~J.} \bibnamefont{Weinberg}},
  \bibinfo{journal}{Phys.Rev.} \textbf{\bibinfo{volume}{D7}},
  \bibinfo{pages}{1888} (\bibinfo{year}{1973}).

\bibitem{Schaefer:2006sr}
\bibinfo{author}{\bibfnamefont{B.-J.} \bibnamefont{Schaefer}} \bibnamefont{and}
  \bibinfo{author}{\bibfnamefont{J.}~\bibnamefont{Wambach}},
  \bibinfo{journal}{Phys.Part.Nucl.} \textbf{\bibinfo{volume}{39}},
  \bibinfo{pages}{1025} (\bibinfo{year}{2008}).

\bibitem{Baym:1982ca}
\bibinfo{author}{\bibfnamefont{G.}~\bibnamefont{Baym}},
  \bibinfo{author}{\bibfnamefont{B.}~\bibnamefont{Friman}}, \bibnamefont{and}
  \bibinfo{author}{\bibfnamefont{G.}~\bibnamefont{Grinstein}},
  \bibinfo{journal}{Nucl.Phys.} \textbf{\bibinfo{volume}{B210}},
  \bibinfo{pages}{193} (\bibinfo{year}{1982}).

\bibitem{Landau1969}
\bibinfo{author}{\bibfnamefont{L.}~\bibnamefont{Landau}} \bibnamefont{and}
  \bibinfo{author}{\bibfnamefont{E.}~\bibnamefont{Lifshitz}},
  \bibinfo{title}{Statistical Physics} (\bibinfo{publisher}{Addison-Wesley,
  Reading, MA}, \bibinfo{year}{1969}).

\end{thebibliography}
\end{document}